\documentclass[a4paper,fleqn,usenatbib]{mnras}

\usepackage{amsmath}	% Advanced maths commands
\usepackage{amssymb}	% Extra maths symbols
\usepackage{newtxtt,newtxmath}
\usepackage[T1]{fontenc}
\usepackage{ae,aecompl}

\usepackage{graphicx}	% Including figure files
\usepackage{epstopdf}
\usepackage{subcaption}
\usepackage{multicol}
\usepackage{multirow}
\usepackage{float}
\usepackage[usenames,dvipsnames,svgnames,table]{xcolor}
\usepackage{enumerate}

\newcommand{\ha}{H$\alpha$}
\newcommand{\hb}{H$\beta+$[O{\sc iii}]}
\newcommand{\oiii}{[O{\sc iii}]}
\newcommand{\oii}{[O{\sc ii}]}

\newcommand{\ewr}{EW$_{\mathrm{rest}}$}
\newcommand{\msol}{M$_\odot$}
\newcommand{\wtheta}{$w(\theta)$}
\newcommand{\ro}{$r_0$}

\def\code#1{\texttt{#1}}

\title[\hb~and \oii~Clustering Properties since $z \sim 5$]{The clustering of \hb~and \oii~emitters since $z \sim 5$: dependencies with line luminosity and stellar mass}
\author[Khostovan et al.]{A.~A.~Khostovan$^{1}$\thanks{NASA Earth and Space Science Fellow}\thanks{E-mail:
akhostov@gmail.com}, D.~Sobral$^{2,3}$, B.~Mobasher$^{1}$, P.~N.~Best$^{4}$, I. Smail$^{5,6}$, \newauthor
J.~Matthee$^{3}$, B.~Darvish$^{7}$, H.~Nayyeri$^{8}$, S.~Hemmati$^{9}$, J.P.~Stott$^{2,10}$\\
$^{1}$Department of Physics \& Astronomy, University of California, Riverside, United States of America\\
$^{2}$Department of Physics, Lancaster University, Lancaster, LA1 4YB, UK \\
$^{3}$Leiden Observatory, Leiden University, PO Box 9513, NL-2300 RA Leiden, the Netherlands\\
$^{4}$SUPA, Institute for Astronomy, Royal Observatory of Edinburgh, Blackford Hill, Edinburgh EH9 3HJ, UK\\
$^{5}$Centre for Extragalactic Astrophysics, Department of Physics, Durham University, Durham DH1 3LE, UK\\
$^{6}$Institute for Computational Cosmology, Durham University, Durham DH1 3LE, UK\\
$^{7}$Cahill Center for Astronomy and Astrophysics, California Institute of Technology, Pasadena, CA 91125, USA\\
$^{8}$Department of Physics \& Astronomy, University of California, Irvine, CA 92697, USA\\
$^{9}$Infrared Processing and Analysis Center, California Institute of Technology, Pasadena, CA 91125, USA\\
$^{10}$Sub-department of Astrophysics, Department of Physics, University of Oxford, Oxford OX1 3RH, UK}

\date{}

\pubyear{2017}

\begin{document}

\label{firstpage}
\pagerange{\pageref{firstpage}--\pageref{lastpage}}
\maketitle

% Abstract of the Paper
\begin{abstract}
We investigate the clustering properties of $\sim 7000$ \hb~and \oii~narrowband-selected emitters at $z \sim 0.8 - 4.7$ from the High-$z$ Emission Line Survey. We find clustering lengths, \ro, of 1.5 -- 4.0 $h^{-1}$ Mpc and minimum dark matter halo masses of $10^{10.7 - 12.1}$ \msol~for our $z = 0.8 - 3.2$ \hb~emitters and \ro$\sim 2.0$ -- $8.3$ $h^{-1}$ Mpc and halo masses of $10^{11.5 - 12.6}$ \msol~for our $z = 1.5 - 4.7$ \oii~emitters. We find \ro~to strongly increase both with increasing line luminosity and redshift. By taking into account the evolution of the characteristic line luminosity, $L^\star(z)$, and using our model predictions of halo mass given \ro, we find a strong, redshift-independent increasing trend between $L/L^\star(z)$ and minimum halo mass. The faintest \hb~emitters are found to reside in $10^{9.5}$ \msol~halos and the brightest emitters in $10^{13.0}$ \msol~halos. For \oii~emitters, the faintest emitters are found in $10^{10.5}$ \msol~halos and the brightest emitters in $10^{12.6}$ \msol~halos. A redshift-independent stellar mass dependency is also observed where the halo mass increases from $10^{11}$ \msol~to $10^{12.5}$ \msol~for stellar masses of $10^{8.5}$ \msol~to $10^{11.5}$ \msol, respectively. We investigate the interdependencies of these trends by repeating our analysis in a $L_\textrm{line}$ -- $M_\textrm{star}$ grid space for our most populated samples (\hb~$z = 0.84$ and \oii~$z = 1.47$) and find that the line luminosity dependency is stronger than the stellar mass dependency on halo mass. For $L > L^\star$ emitters at all epochs, we find a relatively flat trend with halo masses of $10^{12.5 - 13}$ \msol~which may be due to quenching mechanisms in massive halos which is consistent with a transitional halo mass predicted by models.
\end{abstract}


\section{Introduction}
Our current understanding of galaxy formation and evolution implies that galaxies formed hierarchically and inside dark matter halos, such that the baryon clustering traces the underlying dark matter distribution (see \citealt{Benson2010} for a review and references therein). We thus expect a galaxy-halo connection for which the evolving properties of galaxies are tied into the changes of their host halos. A detailed investigation of the dark matter halo properties of galaxies and their evolution is then crucial in setting constraints on current models of galaxy formation. 

Previous theoretical studies have looked into the galaxy-halo connection in several ways. One such method is by using semi-analytical models that identify dark matter halos from {\it N}-body simulations and populating them with galaxies based on analytic relations of the underlying baryon evolution (see \citealt{Baugh2006} and \citealt{Somerville2015} for reviews). Another method is using halo occupation distribution (HOD) models that use probability distributions of how many galaxies reside in halos with a specific mass (see \citealt{Cooray2002} for a review). A similar approach is abundance matching, which works by assigning the most massive galaxies to the most massive halos (e.g., \citealt{Behroozi2010,Guo2010,Moster2010}), although there are several caveats in this technique such as the scatter of stellar mass for a given halo mass and the contribution of satellite galaxies (e.g., \citealt{Contreras2015}). 

On the observational side, large, wide-field, spectroscopic surveys (e.g., SDSS: \citealt{York2000}, 2dFGRS: \citealt{Colless2001}, DEEP2: \citealt{Davis2003}, PRIMUS: \citealt{Coil2011}, GAMA: \citealt{Driver2011}) in the last two decades have made it possible to investigate the clustering properties of galaxies as a function of different types (e.g., colors, luminosities, star formation rates, and stellar masses). For example, studies have found that red, passive galaxies are more clustered than blue, active galaxies (e.g., \citealt{Norberg2002,Zehavi2005,Coil2008,Zehavi2011,Guo2013}). In terms of stellar continuum luminosities (e.g. $B$-band luminosity), there is evidence for a luminosity-dependency with halo mass such that brighter galaxies tend to populate more massive halos (e.g., \citealt{Marulli2013,Guo2014,Harikane2016}).

There are a number of observational studies that have investigated the dependence of clustering strength/dark matter halo mass on stellar mass (e.g., \citealt{Meneux2008,Meneux2009,Wake2011,Lin2012,Mostek2013,McCracken2015}). The connection between dark matter halo and stellar mass also forms the basis of abundance matching (e.g., \citealt{Behroozi2013,Skibba2015,Harikane2016}). However, recent studies have shown this to be more complicated with the relation between the stellar mass and halo mass also being a function of other properties. For example, \citet{Matthee2017} used the hydrodynamical EAGLE simulation to investigate the scatter in the stellar-halo mass relation and came to the conclusion that either the scatter is mass dependent or it depends on more complex halo properties. \citet{Contreras2015} studied the galaxy-halo connection using two independent $N$-body simulations and found a monotonic increasing trend between halo mass and galaxy properties, such as stellar mass, although they find a considerable scatter for a given halo mass. A recent observational study by \citet{Coil2017} using the combined PRIMUS and DEEP2 surveys concluded that there is a wide range of stellar masses for a given halo mass and found that the relationship is also very much dependent on the specific star formation rate.

Other studies have also explored the dependencies on halo mass based on star-formation rates (SFRs) and specific SFRs (sSFRS). Recent measurements using \ha~(tracing the instantaneous SFR) up to $z \sim 2$ find that the clustering signal strongly increases with increasing line luminosity \citep{Sobral2010,Stroe2015,Cochrane2017}. Surprisingly, \citet{Sobral2010} found that the dependency is also redshift-independent in terms of $L/L^\star(z)$, with $L^\star$ being the characteristic \ha~luminosity at each redshift, equivalent to a characteristic SFR (SFR$^\star$, \citealt{Sobral2014}). These studies also find that the trend may flatten for emitters with line luminosities $\gtrsim L^\star$ where emitters seem to reside in $\sim 10^{13 - 13.5}$ \msol~halos. This is consistent with the typical halo masses of AGN-selected samples \citep{Hickox2009,Mendez2016} with recent spectroscopic studies finding that the AGN fraction increases with line luminosity such that emission line-selected galaxies with $L \gg L^\star$ are primarily AGNs \citep{Sobral2016}. \citet{Dolley2014} used a 24\micron-selected sample between $0.2 < z < 1.0$ and found a dependency between total infrared luminosity and halo mass. Using the DEEP2 samples, \citet{Mostek2013} found that the clustering amplitude for $z \sim 1$ blue galaxies strongly increases with SFR and decreasing sSFR while the red population showed no significant correlation with SFR and sSFR. 

The trends highlighted above are based on samples of the nearby Universe and a handful of $z \sim 1 - 2$ studies. When and how these trends formed is important for our understanding of how halos and galaxies coevolve and also helps to constrain galaxy evolution models. In order to effectively study the clustering properties of galaxies, we require samples that are well-defined in terms of selection criteria, cover a range of redshifts to trace the evolving parameters over cosmic time, cover multiple and large comoving volumes to reduce the effects of cosmic variance, span a wide range in physical properties to properly subdivide the samples (e.g., line luminosity bins), and have known redshifts. 

In this study, we use a sample of \hb~and \oii~emission line-selected galaxies from \citet{Khostovan2015} to study the clustering properties and dependencies with line luminosity and stellar mass up to $z \sim 5$ in 4 narrow redshift slices per emission line. Since our samples are emission line-selected, this gives us the advantage of knowing the redshifts of our sources within $\sigma_z = 0.01 - 0.03$ (based on the narrowband filter used) and forms a simple selection function, which is usually not the case with previous clustering studies using either broadband filters or spectroscopic surveys. Our samples are also large enough ($\sim 7000$ sources) to properly subdivide to study the dependency of galaxy properties on the clustering strength and spread over the COSMOS and UDS fields ($\sim 2$ deg$^2$) to reduce the effects of cosmic variance.

This paper is structured as follows: in \S \ref{sec:sample}, we describe our samples and the mock random samples used in the clustering measurements. In \S \ref{sec:clustering} we present our methodology of measuring the angular correlation function, discuss the effects of contamination, describe how we corrected for cosmic variance, present our measurements of the spatial correlation function, and describe our model to convert the clustering length to minimum dark matter halo mass. In \S \ref{sec:clustering_properties} we analyze the results for the full sample measurements in terms of the clustering length and halo masses. In \S \ref{sec:dependencies} we look at the individual dependencies with halo mass starting with line luminosity and followed by stellar mass. We then show the dependency with halo mass in a line luminosity-stellar mass grid space. In \S \ref{sec:discussion} we present our interpretations of the results. We present our main conclusions in \S \ref{sec:conclusions}. 

\begin{figure*}
\centering
% left, lower, right, upper
\includegraphics[width=1.1\textwidth,trim={65 10 15 25},clip]{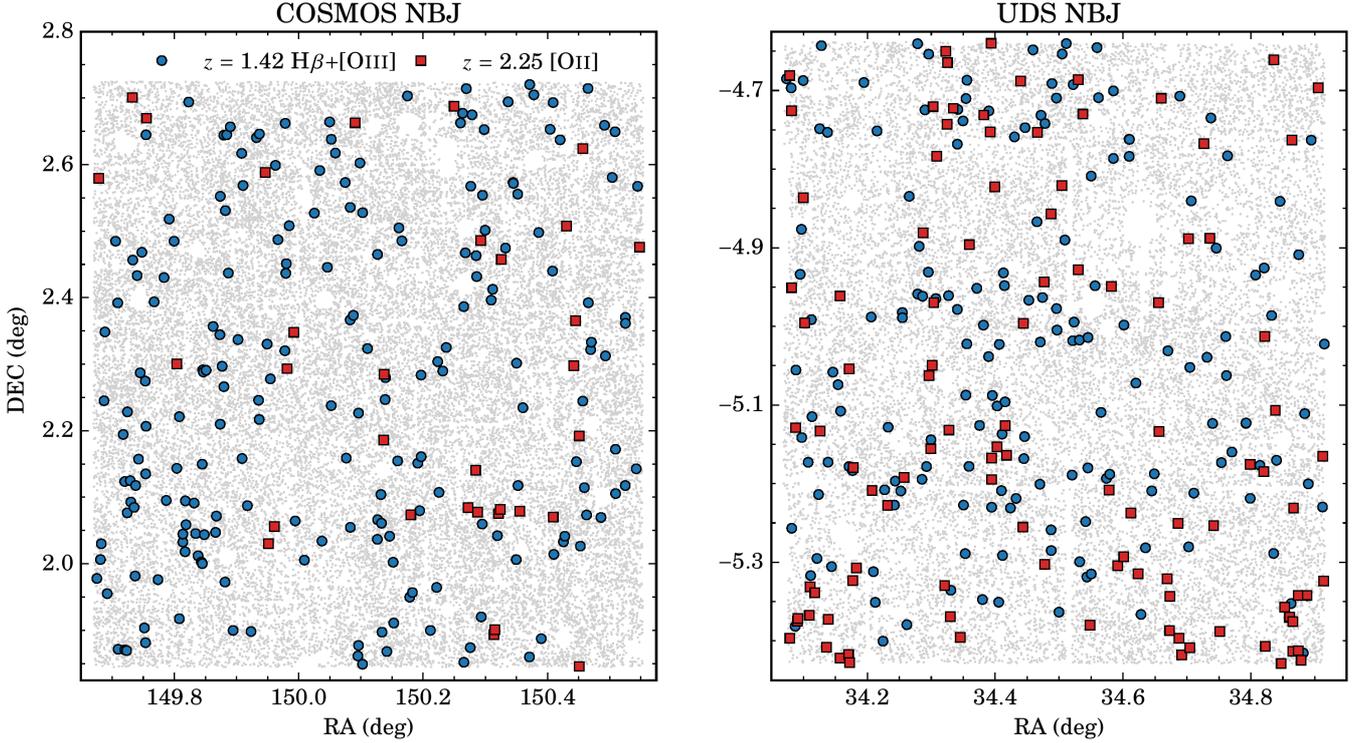}
\caption{The full COSMOS and UDS on-sky coverages with the NBJ filter. Shown in {\it blue circles} and {\it red squares} are the $z = 1.42$ \hb~and $z = 2.25$ \oii~emitters, respectively. The {\it grey dots} are all sources in the raw catalog used to select emission-line galaxies and clearly outline the masked regions which are associated with bright stars and artifacts. We refer the reader to \citet{Sobral2013} for a detailed description of how the masked regions were identified. The spatial distribution shown for both the \hb~and \oii~emitters already shows, visually and qualitatively, signatures of a non-random distribution. To properly quantify the clustering signal, we need to produce random samples that carefully take into account masked regions as outlined above.}
\label{fig:masks}
\end{figure*}

Throughout this paper we assume $\Lambda$CDM cosmology with $H_0 = 70$ km s$^{-1}$, $\Omega_\textrm{m} = 0.3$, and $\Omega_\Lambda = 0.7$. All stellar masses reported assume a Chabrier initial mass function.

\section{Sample}
\label{sec:sample}

\subsection{Emission-Line Galaxy Sample}
In this study, we use the large sample of \hb~and \oii~selected emission-line galaxies from the narrowband High-$z$ Emission Line Survey (HiZELS; \citealt{Geach2008,Sobral2009,Sobral2012,Sobral2013}) presented by \citet{Khostovan2015}. Our samples are distributed over the COSMOS \citep{Scoville2007} and UDS \citep{Lawrence2007} fields with a combined areal coverage of $\sim 2$ deg$^{2}$ which equates to comoving volume coverages of $\sim 10^6$ Mpc$^3$. The sample consists of 3475 \hb~emitters at narrow redshift slices of $z = 0.84$, 1.42, 2.23, and 3.24 and 3298 \oii~emitters at $z = 1.47$, 2.25, 3.34, and 4.69. There are 223 and 219 spectroscopically confirmed \hb~and \oii~emitters, respectively, drawn from the UDSz Survey \citep{Bradshaw2013, McLure2013}, Subaru-FMOS measurements \citep{Stott2013}, Keck/DEIMOS and MOSFIRE measurements (Nayyeri et al., in prep), PRIsm MUlti-object Survey (PRIMUS; \citealt{Coil2011}), and VIMOS Public Extragalactic Redshift Survey (VIPERS; \citealt{Garilli2014}). Recent Keck/MOSFIRE measurements of $z = 1.47 - 3.34$ emitters are also included as well as recent VLT/VIMOS measurements for UDS sources (Khostovan et al., in prep).

The selection criteria used is explained in detail in \citet{Khostovan2015}. In brief, \hb~and \oii~emitters are selected based on a combination of spectroscopic measurements, photometric redshifts, and color-color selections (in order of priority) from the HiZELS narrowband color excess catalog of \citet{Sobral2013}. Sources that have detections in multiple narrowband filters were also included in the final sample as the multiple emission line detections are equivalent to spectroscopic confirmation (e.g., the detection of \oii~in NB921 and \ha~in NBH, see \citealt{Sobral2012}; \oiii~in NBH and \ha~in NBK, \citealt{Suzuki2016}; see also \citealt{Matthee2016} and \citealt{Sobral2017} for dual NB-detections of Ly$\alpha$ and \ha~emitters at $z = 2.23$). 

Stellar masses of the sample were measured by \citet{Khostovan2016} using the SED fitting code of MAGPHYS \citep{Cunha2008}, which works by balancing the stellar and dust components (e.g., the amount of attenuated stellar radiation is accounted for in the infrared). The level of AGN contamination was assessed by \citet{Khostovan2015} to be on the order of $\sim 10 - 20\%$ using the 1.6\micron~bump as a proxy via the color excesses in the {\it Spitzer} IRAC bands. Overall, the sample covers a wide range in physical properties with stellar masses between 10$^{8 - 11.5}$ \msol, \ewr~between $10 - 10000$ \AA, and line luminosities between 10$^{40.5 - 43.0}$ erg s$^{-1}$, providing a wealth of different types of ``active" galaxies (star-forming + AGN; \citealt{Khostovan2016}). This is important when investigating the connection between physical and clustering properties of galaxies.

A unique advantage of narrowband surveys in terms of clustering studies is knowing the redshift distribution of each line (emission line-selected) which removes any redshift projections. Figure \ref{fig:masks} shows the spatial distribution of the NBJ samples (\hb~$z = 1.42$ and \oii~$z \sim 2.25$) where, visually, it is clear that sources in both samples have a non-random, spatial clustering.

\subsection{Random Sample}
\label{sec:random}
When looking for a clustering signal, an equivalent and consistent random catalog is required to test for a non-random spatial distribution within the sample. If all the sources within the sample are consistent with a random spatial distribution, then no spatial correlation would exist within the errors. Therefore, the methodology of creating the random sample has to be consistent with the real dataset in terms of depth, survey geometry, and masked regions (see Figure \ref{fig:masks}).

We create our random samples on an image-by-image basis in order to take into account the different survey depths.\footnotemark As we also want to investigate the dependency with line luminosity and stellar mass (see \S \ref{sec:dependencies}), we populate each image using the line luminosity functions of \citet{Khostovan2015}. For each image, we calculate the total effective area which takes into account the masked areas. We then integrate the \citet{Khostovan2015} luminosity functions down to the $3\sigma$ detection limit of each image to calculate the total number of sources expected within the image area. This is then rescaled up by a factor of $10^5$ such that each random sample generated has a total of $\sim 10^6$ mock sources for each field. Figure \ref{fig:masks} shows the masked regions of the NBJ images for both the COSMOS and UDS fields that are taken into account when generating the random samples.

\footnotetext{Refer to Table 2 of \citet{Sobral2013} for information regarding the depth of each image.}

\section{Measuring the Clustering of \hb~and \oii~emitters}
\label{sec:clustering}

\subsection{Angular Correlation Function}

The two-point angular correlation function (ACF; \wtheta) is defined as:
\begin{eqnarray}
\centering
dP_{12} = \mathcal{N}^2 [1+w(\theta)] d\Omega_1 d\Omega_2
\end{eqnarray}
where $P_{12}$ is the excess probability of finding two galaxies (galaxy 1 and galaxy 2) within a solid angle, $\Omega$, at a given angular separation, $\theta$, and with a mean number density $\mathcal{N}$. Galaxies are randomly distributed for the case of \wtheta$ = 0$ while a non-zero \wtheta~corresponds to a non-random distribution. We use the \citet[LS]{Landy1993} estimator to measure the two-point angular correlation function as it has been shown to be the most reliable and has the best edge corrections when compared to other major estimators \citep{Kerscher2000}. The LS estimator is defined as:
\begin{eqnarray}
\centering
w(\theta) = 1 + \Bigg(\frac{N_R}{N_D}\Bigg)^2 \frac{DD(\theta)}{RR(\theta)} - 2 \frac{N_R}{N_D} \frac{DR(\theta)}{RR(\theta)}
\label{eqn:wtheta}
\end{eqnarray}
where \wtheta~is the angular correlation function, $DD$ is the number of data-data pairs, $RR$ is the number of random-random pairs, $DR$ is the number of data-random pairs, $\theta$ is the angular separation, and $N_R$ and $N_D$ are the total number of random and data sources, respectively. The error associated with the LS estimator is defined as:
\begin{eqnarray}
\centering
\Delta w(\theta) = \frac{1 + w(\theta)}{\sqrt{DD(\theta)}}
\label{eqn:errors}
\end{eqnarray}
which assumes Poisson error.

\begin{figure}
\centering
\includegraphics[width=\columnwidth,trim={5 20 38 50},clip=true]{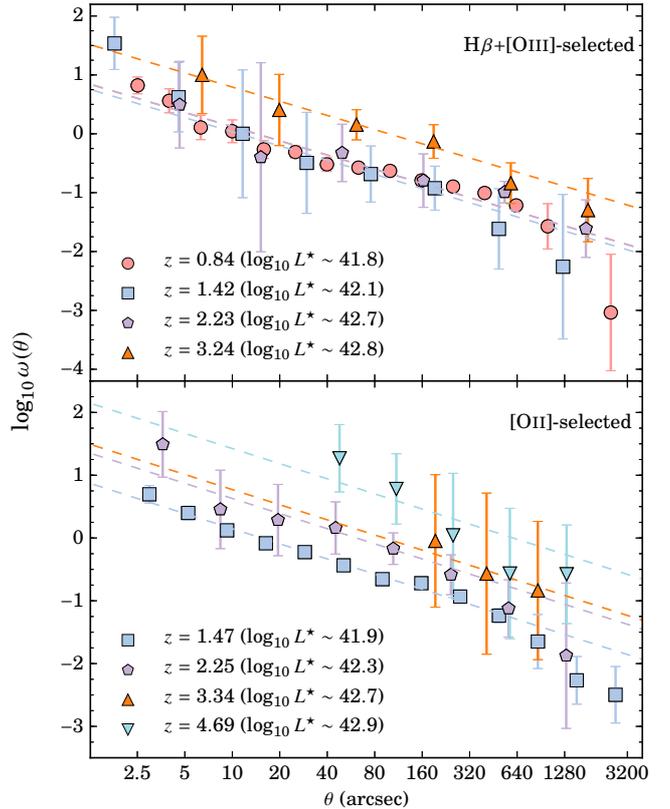}
\caption{The angular correlation function based on the median of all the 2000 realizations per sample with the corresponding Limber approximation fits. All the data points are calculated using the LS estimator. The fits shown are constrained to angular separations for which the ACF is best described as a power law with slope, $\beta = -0.8$. There is evidence for an evolution in the clustering amplitude, but we stress the point that the clustering signal is sensitive to the range of physical properties (e.g., luminosity and stellar mass), which we explore in \S \ref{sec:dependencies}.}
\label{fig:ACF}
\end{figure}

Due to our small sample sizes in comparison to other clustering studies (e.g., SDSS), binning effects could introduce uncertainties in measuring the ACFs. This is basically a signal-to-noise problem where due to the small sample sizes, the way one bins can affect the measured data-data and data-random pairs. For example, bin sizes that are too small will result in bins of data-data pairs (signal) that are not sufficiently populated such that the random-random pairs (noise) will dominate the measured \wtheta. 

To take this into account, we measure the ACF 2000 times assuming Poisson errors as described in Equation \ref{eqn:errors} with varying bin centers and sizes. For each ACF, we apply a random bin size ($\Delta \log \theta = 0.05 - 0.25$ dex) with $\theta_\mathrm{min} = 1.0''$ to $5.0''$ (randomly selected per ACF) and $\theta_\mathrm{max} = 3100''$. Each realization draws 10 - 100 times the number of real sources from the random sample discussed in Section \ref{sec:random} and the number of data-data, random-random, and data-random pairs are measured. We then fit a power law of the form:
\begin{eqnarray}
\centering
w(\theta) &=& A_w \Bigg(\theta^\beta - \textrm{IC} \Bigg) \nonumber \\
IC &=& \frac{\sum{RR \theta^\beta}}{\sum{RR}}
\label{eqn:wtheta}
\end{eqnarray}
with $A_w$ as the clustering amplitude and $\beta$ as the power-law slope. The second equation is the integral constraint (IC; \citealt{Roche2002}) that takes into account the limited survey area. We note that the integral constraint has a marginal effect on our measurements of \ro~as HiZELS coverage is $> 1$ deg$^2$. The final measurements and errors for $A_w$ and the clustering length (\ro; see \S\ref{sec:real_space}) are based on the distributions of values from the 2000 ACFs. In this way, we take into account the effects associated with binning.

Table \ref{table:full_sample_results} shows our $A_w$ and $\beta$ measurements. We find that our measurements are reasonably consistent (within $\sim 1\sigma$) with $\beta \sim -0.80$. We also fit Equation \ref{eqn:wtheta} with a fixed $\beta = -0.80$ (fiducial value in clustering studies) and use these measurements throughout the rest of the paper.

Figure \ref{fig:ACF} shows the median $w(\theta)$ for the 2000 realizations and the fits for the best-fitted $A_w$. We find signs of the $1$-halo term (small-scale clustering/contribution of satellite galaxies) at angular separations $< 20''$ ($\sim 150$ kpc) for the $z = 0.84$ \hb~sample. This is the deepest of all the \hb~samples and probably includes faint, dwarf-like systems that can be potential satellites (the sample includes sources with stellar masses down to $10^{8.5}$ \msol). The deviation from the power law fit seen for the lowest angular separation bin in the $z = 1.42$ \hb~correlation function is consistent with the $1-$halo term, but this is quite weak (within $1\sigma$ deviation). We find no significant detection of the $1-$halo term in the \oii~samples. One possible cause for the $1-$halo term is the presence of large overdense regions that can increase the satellite fraction. For example, there is a $\sim10$ Mpc-scale structure at $z = 0.84$ that contains several X-ray confirmed clusters/groups and large filaments within the COSMOS field (e.g., \citealt{Sobral2011}, \citealt{Darvish2014}) but we defer from a detailed analysis of the satellite fractions as it is beyond the scope of this work.

\subsection{Bootstrapping or Poisson Errors?}
There are three main error estimators that are typically employed in clustering studies: bootstrapping, jackknifing, and Poisson. In the case that Poisson errors are assumed (as is the case with this study), then the errors are defined as shown in equation \ref{eqn:errors}. \citet{Norberg2009} studied these three estimators to see how reliably each measures the `true' errors of the ACFs.  They found that bootstrapping overestimates the errors by $\sim 40$ percent and jackknifing fails at small-scales but can reproduce the errors at large-scales, while Poisson errors were found to underestimate the errors.

One characteristic of the results of \citet{Norberg2009} is that the sample size used in their simulations is comparable to that of SDSS ($10^{5 - 6}$ sources). Since Poisson errors become significantly smaller for larger sample sizes, it then would become apparent that Poisson errors could severely underestimate the `true' errors of the ACFs. This may not be entirely true for our sample sizes, which are typically between $10^{2-3}$ sources. We test this by using our $z = 3.24$ \hb~sample (179 sources) for which the bin size and centers were fixed and calculated the ACFs assuming Poisson errors and also bootstrapping with 2000 realizations. We find that the errors on average are similar such that Poisson errors for small sample sizes are comparable to bootstrapping errors. Note that, as described in \S \ref{sec:clustering}, we assume Poisson errors for each individual ACF but also take into account binning effects by repeating our measurements of the ACF with varying bin sizes and centers such that our final measurements are based on the distributions of these realizations.

\subsection{Effects of Contamination}
The issue of contamination can be marginal or quite significant and is based on many factors such as the sample selection. Clustering studies typically consider the contaminants in a sample to be randomly distributed, such that the clustering amplitude is underestimated by a factor of $(1 - f)^2$, with $f$ being the contamination fraction. For the clustering length, \ro, this results in an underestimation by a factor of $(1-f)^{2/|\gamma|}$.

The level of contamination was briefly investigated in \citet{Khostovan2015} and was found to be on the order of $\sim 10$ percent for the lowest redshift samples. This would result in a 23 percent increase in $A_w$ and a 12 percent increase in \ro. Note that this assumes that the contaminants are randomly distributed and, hence, lowers the clustering strength, which may not be true for narrowband surveys. For our samples, contaminants could be due to galaxies with misidentified emission lines. For example, a source at $z = 1.47$ that is misidentified as \oii~in the NB921 filter could actually be a $z = 0.84$ \oiii~emitter or a $z = 0.40$ \ha~emitter. Because galaxies selected by nebular emission lines are shown to be clustered as well (see below and \citealt{Sobral2010} and \citealt{Cochrane2017} for \ha), the effects could possibly be negligible and not follow the typical $(1-f)^2$ correction factor. Therefore, we do not correct our measurements due to contamination.

\subsection{Cosmic Variance}
\label{sec:cosmic_variance}
Cosmic variance can greatly affect the clustering measurements. If the areal coverage is small ($\lesssim$ arcmin$^2$ scales), then the measured clustering amplitude and subsequent results can vary considerably, especially if the region probed is a significant overdense region or a void. Therefore, it is important that the clustering measurements are done on large fields ($\gtrsim 1$ deg$^2$). 

\citet{Sobral2010} measured the effects of cosmic variance for the HiZELS \ha~$z = 0.84$ sample (734 emitters) on the clustering amplitude. This was done by measuring $A_w$ (fixed $\beta = -0.80$) for randomly sized regions between 0.05 deg$^2$ to 0.5 deg$^2$ with the larger areas randomly sampled 100 times (0.3 - 0.5 deg$^2$) and the smaller areas randomly sampled 1000 times. We refer the reader to Figure 3 of \citet{Sobral2010} where they show that the uncertainty in $A_w$ (in percentage) is related to the area covered and is best fit with a power-law of the form $20 \times \Omega^{-0.35}$, with $\Omega$ representing the area in units of deg$^2$.

We note that the HiZELS coverage at the time of \citet{Sobral2010} was only 1.3 deg$^2$ in the COSMOS field and used only $J$-band coverage. In this paper we are using the current HiZELS coverage (all four narrowband filters in $zJHK$) which includes both the COSMOS and UDS fields for a combined areal coverage of $\sim 2$ deg$^2$ \citep{Sobral2013}. This corresponds to a decreased uncertainty of $\sim 16\%$ due to cosmic variance in the measurement of $A_w$. We incorporate this uncertainty by adding $\sim 16\%$ of $A_w$ in quadrature to the error from the fit. For the clustering length, \ro, we propagate the error from $A_w$ and find that the error in \ro~is increased by $\sim 11\%$.

\begin{table*}
\centering
\caption{The clustering properties for our \hb~and \oii~samples. The power-law slope, $\beta$, in the ACF is shown and corresponds to the clustering amplitude, $A_{w,\mathrm{free}}$, which corresponds to when $\beta$ is a free-parameter in the fit. All other measurements shown have $\beta$ fixed to $-0.8$, which corresponds to $\gamma = -1.8$ in the real-space two-point correlation function. $r_{0,\mathrm{exact}}$ is the clustering length measured using the exact Limber equation as defined in Equation \ref{eqn:exact}. Dark matter halo masses are measured using our $r_0$-halo mass models.}
\begin{tabular}{ccccccc}
\hline
\multicolumn{7}{c}{Clustering Properties for Full Sample}\\
\hline
$z$ & $N_D$ & $\beta$ & $A_{w,\mathrm{free}}$ & $A_{w,\mathrm{\beta = -0.8}}$ & $r_{0,\mathrm{exact}}$ & $\log_{10}$ M$_\textrm{min}$\\
& & & (arcsec) & (arcsec) & (Mpc $h^{-1}$) & (\msol~$h^{-1}$) \\
\hline
\multicolumn{7}{c}{\hb~Emitters}\\
\hline
0.84 & 2477 & -0.69$_{-0.03}^{+0.03}$ & 5.19$_{-1.22}^{+1.32}$ & 11.53$_{-2.33}^{+2.33}$ & 1.71$_{-0.19}^{+0.19}$ & 11.18$_{-0.33}^{+0.33}$\\
1.42 & 371 & -0.79$_{-0.04}^{+0.07}$ & 7.47$_{-3.24}^{+3.58}$ & 8.32$_{-2.08}^{+2.18}$ & 1.45$_{-0.20}^{+0.20}$ & 10.70$_{-0.40}^{+0.40}$\\
2.23 & 270 & -0.81$_{-0.12}^{+0.15}$ & 11.10$_{-6.57}^{+12.42}$ & 10.42$_{-2.62}^{+2.80}$ & 2.43$_{-0.31}^{+0.31}$ & 11.61$_{-0.22}^{+0.22}$\\
3.24 & 179 & -0.78$_{-0.03}^{+0.04}$ & 42.28$_{-13.56}^{+13.22}$ & 48.70$_{-10.83}^{+10.71}$ & 4.01$_{-0.49}^{+0.49}$ & 12.08$_{-0.17}^{+0.17}$\\
\hline
\multicolumn{7}{c}{\oii~Emitters}\\
\hline
1.47 & 3285 & -0.83$_{-0.04}^{+0.02}$ & 10.06$_{-2.21}^{+2.66}$ & 11.61$_{-2.34}^{+2.34}$ & 1.99$_{-0.22}^{+0.22}$ & 11.46$_{-0.24}^{+0.23}$\\
2.25 & 137 & -0.78$_{-0.03}^{+0.05}$ & 25.51$_{-9.18}^{+9.08}$ & 29.99$_{-7.00}^{+7.24}$ & 3.14$_{-0.41}^{+0.43}$ & 12.03$_{-0.20}^{+0.21}$\\
3.34 & 35 & -0.79$_{-0.06}^{+0.23}$ & 53.67$_{-44.95}^{+41.66}$ & 57.49$_{-24.67}^{+22.49}$ & 5.06$_{-0.94}^{+1.08}$ & 12.37$_{-0.24}^{+0.28}$\\
4.69 & 18 & -0.83$_{-0.04}^{+0.04}$ & 208.50$_{-91.58}^{+116.82}$ & 139.44$_{-44.63}^{+53.69}$ & 8.25$_{-1.44}^{+1.54}$ & 12.62$_{-0.20}^{+0.22}$\\
\hline
\end{tabular}
\label{table:full_sample_results}
\end{table*}

\subsection{Real-Space Correlation}
\label{sec:real_space}
The two-point (real-space) correlation function is a useful tool in measuring the physical clustering of galaxies and is best described, empirically, by $\xi = (r/r_0)^\gamma$, with $r_0$ being the clustering length. One key requirement in measuring the two-point correlation function is the redshift distribution of the sample. The benefit of narrowband surveys is that the redshift distribution of the sample is easily derived from the narrowband filter profile (e.g., \citealt{Sobral2010,Stroe2015}) such that it is equivalent to taking a narrow redshift slice of $\sigma_z \sim 0.01 - 0.03$ (depending on the central redshift; see Table 2 in \citealt{Khostovan2015}). 

Traditionally, the Limber approximation \citep{Limber1953} is used to relate the real-space correlation to the angular correlation function. \citet{Simon2007} showed that the approximation works for surveys that use broad filters and for small angular separations but fails for narrow filters and large angular separations. They find that for large angular separations and very narrow filters, $\omega(\theta)$ becomes a rescaled version of $\xi(r)$ where the slope of \wtheta~changes from $\gamma+1$ to $\gamma$. \citet{Sobral2010} used the exact Limber equation proposed by \citet{Simon2007} and found that, for their sample of $z = 0.84$ \ha~emitters, the break-down in the Limber approximation occurs at angular separations $\sim 600''$ with an $r_0 = 2.6\pm0.3 h^{-1}$ Mpc measured from the approximation and $r_0 = 2.7\pm0.3 h^{-1}$ Mpc from the exact equation. 

\begin{figure}
\centering
\includegraphics[width=\columnwidth,trim={0 0 0 30},clip=true]{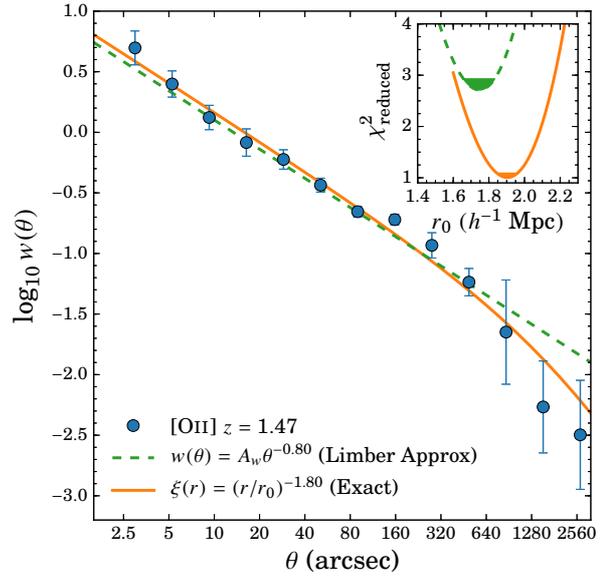}
\caption{The angular correlation function for the $z = 1.47$ \oii~sample. Shown are the observed \wtheta~measurements as in Figure \ref{fig:ACF} with the corresponding Limber approximation and exact Limber equation fits. We use the full range of angular separations for both fits, even though the Limber approximation is found to fail at $\theta \sim 500''$. The exact equation results in a reduced $\chi^2 \approx 1$ compared to $\approx 2.8$ when using the Limber approximation. The clustering lengths are \ro$_\mathrm{,exact} = 1.90\pm0.21$ compared to \ro$_{\mathrm{,limber}} = 1.75\pm0.21$ (errors corrected for cosmic variance). The errors shown in the $\chi^2$ distribution are only based on the fits. The results shown here signify the importance of the exact Limber equation when using narrowband samples for large angular separations.}
\label{fig:OII_ACF}
\end{figure}

We adopt the exact equation presented by \citet{Simon2007} and used by \citet{Sobral2010} to relate the real-space and angular correlation functions and calculate $r_0$. The relation is described as:
\begin{eqnarray}
\omega(\theta) &=& \frac{r_0^{-\gamma}}{1+\cos{\theta}} \int\limits_0^\infty \int\limits_{\bar{r}\sqrt{2(1-\cos{\theta})}}^{2\bar{r}} \frac{2 p(\bar{r} - \Delta) p(\bar{r} + \Delta)}{R^{-\gamma-1} \Delta} \mathrm{d}R \mathrm{d}\bar{r} \nonumber\\
\Delta &=& \sqrt{\frac{R^2 - 2 \bar{r}^2 (1 - \cos{\theta})}{2 (1 + \cos{\theta})}}
\label{eqn:exact}
\end{eqnarray}
where $p$ is the filter profile in radial comoving distance, which is written as the mean spatial position of two sources, $r_1$ and $r_2$, such that $\bar{r} = (r_1 + r_2)/2$ with $R$ being the distance between the two sources using the law of cosines. We refer the reader to \citet{Simon2007} for a detailed description regarding the derivation of this equation. The filter profile, which traces the underlying redshift distribution of the sample, is assumed to be a Gaussian function. We fit for the true filter profile based on the transmission curves of the actual narrowband filters. Table \ref{table:gauss_filter_params} shows a comparison between the properties of the Gaussian and true filters in terms of redshifts. The power law slope of the spatial correlation function is also shown in Equation \ref{eqn:exact} and is assumed to be $\gamma = -1.8$ ($\gamma = \beta - 1$). We use Equation \ref{eqn:exact} to fit \ro~to our measurements of \wtheta.

Figure \ref{fig:OII_ACF} shows the comparison between the Limber approximation (assuming a single power law to describe $w(\theta)$ as shown in Equation \ref{eqn:wtheta}) and the exact Limber equation as described in Equation \ref{eqn:exact} for the $z = 1.47$ \oii~sample. We find that the Limber approximation breaks down at angular separations of $\sim 500''$. As discussed in \citet{Simon2007} and in Appendix \ref{sec:limber_fails}, the point for where the Limber approximation fails is dependent on the filter width (the width of the redshift distribution) and the transverse distance (central redshift). 

Also shown on Figure \ref{fig:OII_ACF} is the reduced $\chi^2$ measurements of the fits. We find that the exact equation has a reduced $\chi^2$ of $\approx 1$ in comparison to $2.8$ for the Limber approximation-based fit with \ro$_\mathrm{,exact} = 1.90\pm0.21$ $h^{-1}$ Mpc compared to \ro$_{\mathrm{,limber}} = 1.75\pm0.21$ $h^{-1}$ Mpc (errors include cosmic variance contribution; see \S \ref{sec:cosmic_variance}). Although both methods produce measurements that are consistent within $1\sigma$ (errors dominated by cosmic variance), our results shown on Figure \ref{fig:OII_ACF} highlights the importance of using the exact Limber equation to measure the clustering length since it can compensate for the rescaling of the ACF due to the effects of using narrowband filters. Throughout the rest of this paper, we refer to $r_0$ as the clustering length measured using Equation \ref{eqn:exact}.

\subsection{Clustering Length to Dark Matter Halo Mass}
\label{sec:dmh_model}
Our theoretical understanding of galaxy formation is that galaxies form with the assistance of the gravitational potentials of dark matter halos such that all galaxies reside in a halo. In effect, the spatial clustering of galaxies is then related to the clustering of dark matter. \citet{Matarrese1997} and \citet{Moscardini1998} used this link between galaxies and dark matter halos to predict the clustering length of a sample for a given minimum dark matter halo mass and redshift. In this section, we use the same methodology used to generate their predictions, but update to the latest cosmological prescriptions.

We first begin by measuring the matter-matter spatial correlation function using a suite of cosmological codes called \code{Colossus} \citep{Diemer2015}. This is calculated by taking the Fourier transform of the matter power spectrum, assuming an \citet{Eisenstein1998} transfer function. We then calculate the effective bias by using the following equation:
\begin{eqnarray}
b_{eff}(z) = \frac{\int_{\textrm{M}_\mathrm{min}}^\infty b_h(M,z) n(M,z) \mathrm{d}M}{\int_{\textrm{M}_\mathrm{min}}^\infty n(M,z) \mathrm{d}M}
\label{eqn:halo_bias}
\end{eqnarray}
where $b_h(M,z)$ and $n(M,z)$ are the halo bias and mass functions, respectively. The effective bias is defined as the integrated halo bias and mass functions above some minimum dark matter halo mass, \textrm{M}$_\mathrm{min}$, and normalized to the number density of halos. We then relate the effective bias to the spatial correlation of galaxies by:
\begin{eqnarray}
b^2_{eff} = \xi_{gg}/\xi_{mm}
\end{eqnarray}
with $\xi_{gg}$ and $\xi_{mm}$ being the galaxy-galaxy and matter-matter spatial correlation functions, respectively.

We use the \citet{Tinker2010} halo bias prescription and the \citet{Tinker2008} halo mass function. The previous predictions of \citet{Matarrese1997} and \citet{Moscardini1998} used the \citet{Press1974} halo mass function and \citet{Mo1996} halo bias functions. Their assumed $\Lambda$CDM cosmology was also different ($H_0 = 65$ km s$^{-1}$ Mpc$^{-1}$, $\Omega_m = 0.4$, and $\Omega_\Lambda = 0.6$) than the current measurements. We present a discussion regarding the uncertainties of assuming a bias and mass function in Appendix \ref{sec:halo_uncertain}.

Note that our approach is very much similar to the methodology used in halo occupation distribution (HOD) modeling (e.g., \citealt{Kravtsov2004}). In comparison to the framework of HOD, we are assuming that all galaxies are centrals (only one galaxy occupies each host halo) and reside in halos with mass $\gtrsim \textrm{M}_\textrm{min}$. This is an oversimplification in comparison to typical HOD models where we have only one free parameter (minimum dark matter halo mass), but we note that HOD modeling typically employs 3 - 5 free parameters (e.g., \citealt{Kravtsov2004,Zheng2005}) with even more complex models incorporating 8 free parameters (e.g., \citealt{Geach2012}). We instead resort to using our one parameter approach but caution the reader that directly comparing our results with minimum halo masses is inconsistent. Any study from the literature that is used to compare with our results in this paper have their minimum halo masses computed using their \ro~measurements and our \ro-halo mass model.

\section{Clustering of \hb~and \oii~Emitters}
\label{sec:clustering_properties}

\begin{figure}
\centering
\includegraphics[width=\columnwidth,trim={20 1 53 30}, clip=true]{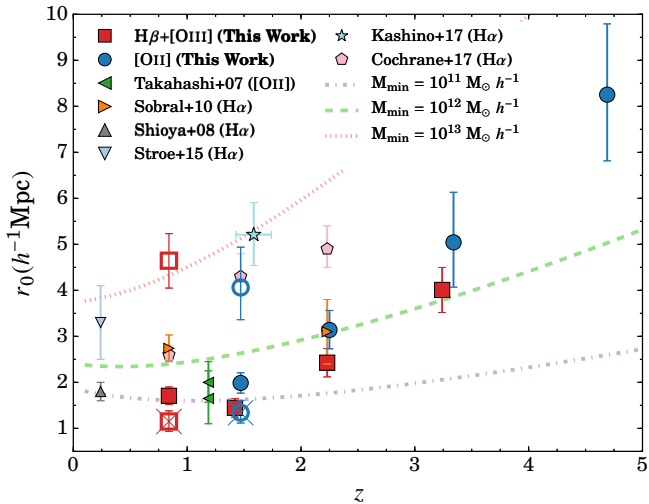}
\caption{Shown is the evolution of the clustering length up to $z \sim 5$. Included are the predicted clustering lengths for minimum dark matter halo masses between $10^{11-13}$ \msol. Although there is a clear sign of a redshift evolution in \ro, we stress the point that this is due to selection bias such that these measurements are sensitive to the range of physical properties, such as line luminosity. As a demonstration, we overlay the brightest ({\it open symbol}) and faintest ({\it open symbol with a cross}) line luminosity bins (see Table \ref{table:HB_sample_results}) with the symbol type and color consistent with that used for the full sample measurement. The brightest emitters are found to have $r_0$ measurements $\sim 2 - 3$ times that of the full sample and the faintest emitters with $\sim 50\%$ lower $r_0$ values.}
\label{fig:clustering_full_sample}
\end{figure}

Figure \ref{fig:clustering_full_sample} shows the evolution of $r_0$ for \hb~and \oii~emitters up to $z \sim 3$ and $\sim 5$, respectively. These are the first measurements of the clustering length for \hb~and \oii~emission-line galaxies to be reported. Included are the \ro~predictions for dark matter halos with minimum masses between $10^{11} - 10^{13}$ \msol~based on our model described in \S \ref{sec:dmh_model}.

We find that, based on the full population of emitters in each sample, \hb~emitters tend to reside in $\sim 10^{10.7} - 10^{12.1}$ \msol~dark matter halos while the \oii~emitters are found to vary less with $\sim 10^{11.5}$ \msol~at $z = 1.47$ to $\sim 10^{12.6}$ \msol~at $z = 4.69$, although these are driven by selection effects (e.g., highest redshift sample will be biased towards higher line luminosities which, as shown in \S \ref{sec:line_luminosity}, leads to higher \ro). In comparison to each other, all overlapping samples, except for the $z = 1.47$ samples, have similar $r_0$ measurements within $1 \sigma$ error bars. This then suggests that \hb- and \oii-selected galaxies reside in dark matter halos with similar masses. Included in Figure \ref{fig:clustering_full_sample} are the \ha~measurements of \citet{Shioya2008}, \citet{Sobral2010}, \citet{Stroe2015}, \citet{Cochrane2017}, and \citet{Kashino2017}. The \citet{Sobral2010} measurement at $z = 2.23$ is consistent with that of the \hb~and \oii~samples at the same redshift, suggesting that \hb- and \oii-selected emitters reside in dark matter halos with similar masses as \ha-selected emitters and can be tracing a similar underlying population of star-forming/active galaxies. We also include the $z \sim 1.2$ \oii~measurements of \citet{Takahashi2007}. Although our closest sample in terms of redshift is at $z = 1.47$, we find that our measurements are in agreement.

Despite the agreement between \ha, \hb, and \oii~samples, we note that such a comparison is not entirely fair. An example is the \ha~measurement of \citet{Stroe2015} and \citet{Shioya2008}. Both cover the same redshift range of $z = 0.24$, but the \citet{Shioya2008} has a depth of $\sim 10^{39.5}$ erg s$^{-1}$ in $L_{\textrm{\ha}}$ while the \citet{Stroe2015} depth is $\sim 10^{41.0}$ erg s$^{-1}$ and covers significantly larger volumes. This results in a factor of two difference in the $r_0$ measured and almost two orders of magnitude difference in the minimum dark matter halo mass by these two studies which arises from the dependency of the clustering length with line luminosity (see \S \ref{sec:line_luminosity}). 

As a demonstration of this same feature, we show \ro~of the brightest (open symbols) and faintest (open symbols with a cross) galaxies in our \hb~$z = 0.84$ and \oii~$z = 1.47$ samples. We find that the most luminous (faintest) galaxies have higher (lower) clustering lengths relative to the full sample measurement. This suggests a line luminosity dependency not just in the \ha~measurements, but also in the \hb~and \oii~measurements. Therefore, any comparison, as shown in Figure \ref{fig:clustering_full_sample}, needs to be interpreted with caution as each measurement for a full sample will be dependent on how wide a range of line luminosities is covered. For example, the $r_0$ measured for the $z = 4.69$ \oii~sample is biased towards higher $r_0$ values since the sample is biased towards the brightest \oii~emitters. To investigate the redshift evolution of the clustering and dark matter halo properties of galaxies, we need to then study its dependencies.

\begin{table}
\centering
\caption{Clustering Properties of the \hb~samples as a function of line luminosities and stellar masses. We include $L^\star(z)$ for each sample as measured by \citet{Khostovan2015}. All measurements assume a fixed $\gamma = -1.8$. The minimum dark matter halo masses are measured from the \ro~measurements in conjunction with our \ro-halo mass models. All measurements are corrected for cosmic variance by adding in quadrature 11\% of \ro~in the total error cited.}
\resizebox{\columnwidth}{!}{%
\begin{tabular}{cccc}
\hline
Subsample & $N_D$ & $r_{0,\mathrm{exact}}$ & $\log_{10}$ M$_\textrm{min}$\\
& & (Mpc $h^{-1}$) & (\msol~$h^{-1}$) \\
\hline
\multicolumn{4}{c}{\hb~$z = 0.84$ ($\log_{10} L^\star = 41.79^{+0.03}_{-0.05}$)}\\
\hline
$40.50 < \log_{10} L_{\textrm{line}} < 40.60$ & 188 & 1.15$_{-0.22}^{+0.23}$ & 9.48$_{-1.30}^{+1.40}$\\
$40.60 < \log_{10} L_{\textrm{line}} < 40.70$ & 175 & 1.46$_{-0.22}^{+0.23}$ & 10.66$_{-0.57}^{+0.59}$\\
$40.70 < \log_{10} L_{\textrm{line}} < 40.80$ & 150 & 1.46$_{-0.25}^{+0.26}$ & 10.67$_{-0.63}^{+0.67}$\\
$40.80 < \log_{10} L_{\textrm{line}} < 41.00$ & 279 & 1.46$_{-0.21}^{+0.20}$ & 10.67$_{-0.52}^{+0.52}$\\
$41.00 < \log_{10} L_{\textrm{line}} < 41.15$ & 538 & 1.77$_{-0.21}^{+0.22}$ & 11.28$_{-0.34}^{+0.34}$\\
$41.15 < \log_{10} L_{\textrm{line}} < 41.30$ & 404 & 1.89$_{-0.23}^{+0.23}$ & 11.46$_{-0.31}^{+0.31}$\\
$41.30 < \log_{10} L_{\textrm{line}} < 41.60$ & 492 & 2.08$_{-0.24}^{+0.25}$ & 11.69$_{-0.27}^{+0.28}$\\
$41.60 < \log_{10} L_{\textrm{line}} < 41.80$ & 131 & 3.18$_{-0.42}^{+0.44}$ & 12.53$_{-0.22}^{+0.23}$\\
$41.80 < \log_{10} L_{\textrm{line}} < 41.95$ & 51 & 3.24$_{-0.46}^{+0.51}$ & 12.55$_{-0.24}^{+0.26}$\\
$41.95 < \log_{10} L_{\textrm{line}} < 42.55$ & 61 & 4.64$_{-0.60}^{+0.59}$ & 13.10$_{-0.18}^{+0.17}$\\
 & & & \\
$8.50 < \log_{10} M < 8.75$ & 368 & 1.60$_{-0.21}^{+0.22}$ & 11.15$_{-0.32}^{+0.32}$\\
$8.75 < \log_{10} M < 9.00$ & 483 & 1.75$_{-0.22}^{+0.22}$ & 11.35$_{-0.28}^{+0.28}$\\
$9.00 < \log_{10} M < 9.20$ & 391 & 1.74$_{-0.22}^{+0.21}$ & 11.33$_{-0.28}^{+0.27}$\\
$9.20 < \log_{10} M < 9.40$ & 294 & 2.26$_{-0.28}^{+0.28}$ & 11.89$_{-0.26}^{+0.26}$\\
$9.40 < \log_{10} M < 9.70$ & 271 & 2.34$_{-0.29}^{+0.30}$ & 11.96$_{-0.26}^{+0.26}$\\
$9.70 < \log_{10} M < 10.64$ & 213 & 2.56$_{-0.32}^{+0.32}$ & 12.11$_{-0.19}^{+0.20}$\\
$10.64 < \log_{10} M < 11.55$ & 74 & 3.41$_{-0.49}^{+0.46}$ & 12.55$_{-0.22}^{+0.21}$\\
\hline
\multicolumn{4}{c}{\hb~$z = 1.42$ ($\log_{10} L^\star = 42.06^{+0.06}_{-0.05}$)}\\
\hline
$41.92 < \log_{10} L_{\textrm{line}} < 42.02$ & 191 & 1.54$_{-0.25}^{+0.28}$ & 10.87$_{-0.44}^{+0.49}$\\
$42.02 < \log_{10} L_{\textrm{line}} < 42.06$ & 63 & 2.33$_{-0.48}^{+0.49}$ & 11.79$_{-0.39}^{+0.40}$\\
$42.06 < \log_{10} L_{\textrm{line}} < 42.16$ & 58 & 4.30$_{-0.68}^{+0.67}$ & 12.78$_{-0.22}^{+0.22}$\\
$42.16 < \log_{10} L_{\textrm{line}} < 42.26$ & 25 & 4.28$_{-1.08}^{+1.12}$ & 12.78$_{-0.35}^{+0.36}$\\
$42.26 < \log_{10} L_{\textrm{line}} < 42.80$ & 34 & 3.97$_{-0.84}^{+0.82}$ & 12.67$_{-0.31}^{+0.30}$\\
& & & \\
$9.00 < \log_{10} M < 9.50$ & 96 & 2.10$_{-0.36}^{+0.38}$ & 11.54$_{-0.31}^{+0.33}$\\
$9.50 < \log_{10} M < 10.00$ & 99 & 3.00$_{-0.41}^{+0.45}$ & 12.14$_{-0.19}^{+0.21}$\\
$10.00 < \log_{10} M < 10.50$ & 60 & 2.93$_{-0.55}^{+0.66}$ & 12.11$_{-0.26}^{+0.31}$\\
$10.50 < \log_{10} M < 11.00$ & 53 & 3.06$_{-0.55}^{+0.62}$ & 12.18$_{-0.25}^{+0.28}$\\
\hline
\multicolumn{4}{c}{\hb~$z = 2.23$ ($\log_{10} L^\star = 42.66^{+0.13}_{-0.13}$)}\\
\hline
$42.30 < \log_{10} L_{\textrm{line}} < 42.66$ & 136 & 2.66$_{-0.44}^{+0.44}$ & 11.77$_{-0.28}^{+0.28}$\\
$42.66 < \log_{10} L_{\textrm{line}} < 42.74$ & 56 & 5.15$_{-0.68}^{+0.64}$ & 12.74$_{-0.17}^{+0.16}$\\
$42.74 < \log_{10} L_{\textrm{line}} < 43.10$ & 57 & 7.38$_{-0.90}^{+0.88}$ & 13.17$_{-0.14}^{+0.14}$\\
& & & \\
$9.25 < \log_{10} M < 10.00$ & 120 & 3.08$_{-0.45}^{+0.47}$ & 11.89$_{-0.25}^{+0.26}$\\
$10.00 < \log_{10} M < 10.50$ & 66 & 3.22$_{-0.50}^{+0.50}$ & 11.97$_{-0.27}^{+0.27}$\\
$10.50 < \log_{10} M < 11.00$ & 41 & 3.48$_{-0.90}^{+0.91}$ & 12.08$_{-0.34}^{+0.35}$\\
\hline
\multicolumn{4}{c}{\hb~$z = 3.24$ ($\log_{10} L^\star = 42.83^{+0.19}_{-0.17}$)}\\
\hline
$42.30 < \log_{10} L_{\textrm{line}} < 42.67$ & 68 & 3.24$_{-0.53}^{+0.51}$ & 11.77$_{-0.25}^{+0.24}$\\
$42.67 < \log_{10} L_{\textrm{line}} < 42.83$ & 67 & 5.56$_{-0.73}^{+0.74}$ & 12.52$_{-0.17}^{+0.17}$\\
$42.83 < \log_{10} L_{\textrm{line}} < 43.18$ & 44 & 6.98$_{-1.00}^{+1.12}$ & 12.80$_{-0.17}^{+0.19}$\\
& & & \\
$9.20 < \log_{10} M < 9.70$ & 56 & 5.09$_{-0.63}^{+0.69}$ & 12.29$_{-0.17}^{+0.18}$\\
$9.70 < \log_{10} M < 10.30$ & 80 & 4.35$_{-0.55}^{+0.65}$ & 12.08$_{-0.17}^{+0.20}$\\
$10.30 < \log_{10} M < 11.00$ & 29 & 5.02$_{-1.04}^{+1.21}$ & 12.27$_{-0.28}^{+0.32}$\\
\hline
\end{tabular}}
\label{table:HB_sample_results}
\end{table}

\begin{table}
\centering
\caption{The clustering properties of \oii~as a function of line luminosity and stellar mass. Table description is the same as that of Table \ref{table:HB_sample_results}. The $z = 3.34$ and $4.69$ measurements are not included in this table as the sample sizes were too small to divide in line luminosity and stellar mass bins. The measurements corresponding to the full samples are shown in Table \ref{table:full_sample_results}.}
\resizebox{\columnwidth}{!}{%
\begin{tabular}{cccc}
\hline
Subsample & $N_D$ & $r_{0,\mathrm{exact}}$ & $\log_{10}$ M$_\textrm{min}$\\
& & (Mpc $h^{-1}$) & (\msol~$h^{-1}$) \\
\hline
\multicolumn{4}{c}{\oii~$z = 1.47$ ($\log_{10} L^\star = 41.86^{+0.03}_{-0.03}$)}\\
\hline
$41.05 < \log_{10} L_{\textrm{line}} < 41.15$ & 200 & 1.34$_{-0.22}^{+0.27}$ & 10.47$_{-0.51}^{+0.61}$\\
$41.15 < \log_{10} L_{\textrm{line}} < 41.25$ & 501 & 1.41$_{-0.18}^{+0.18}$ & 10.62$_{-0.36}^{+0.37}$\\
$41.25 < \log_{10} L_{\textrm{line}} < 41.45$ & 761 & 1.74$_{-0.20}^{+0.20}$ & 11.16$_{-0.28}^{+0.27}$\\
$41.45 < \log_{10} L_{\textrm{line}} < 41.65$ & 638 & 2.47$_{-0.29}^{+0.29}$ & 11.89$_{-0.22}^{+0.21}$\\
$41.65 < \log_{10} L_{\textrm{line}} < 41.85$ & 667 & 2.76$_{-0.32}^{+0.32}$ & 12.08$_{-0.20}^{+0.20}$\\
$41.85 < \log_{10} L_{\textrm{line}} < 42.00$ & 292 & 3.34$_{-0.40}^{+0.40}$ & 12.39$_{-0.19}^{+0.19}$\\
$42.00 < \log_{10} L_{\textrm{line}} < 42.10$ & 101 & 3.23$_{-0.49}^{+0.46}$ & 12.34$_{-0.24}^{+0.23}$\\
$42.10 < \log_{10} L_{\textrm{line}} < 42.20$ & 68 & 3.32$_{-0.50}^{+0.49}$ & 12.38$_{-0.24}^{+0.23}$\\
$42.20 < \log_{10} L_{\textrm{line}} < 42.60$ & 56 & 4.06$_{-0.70}^{+0.88}$ & 12.68$_{-0.25}^{+0.31}$\\
 & & & \\
$8.40 < \log_{10} M < 8.80$ & 217 & 1.51$_{-0.23}^{+0.23}$ & 10.84$_{-0.49}^{+0.49}$\\
$8.80 < \log_{10} M < 9.20$ & 671 & 2.04$_{-0.24}^{+0.24}$ & 11.48$_{-0.21}^{+0.21}$\\
$9.20 < \log_{10} M < 9.40$ & 429 & 1.88$_{-0.24}^{+0.24}$ & 11.33$_{-0.23}^{+0.24}$\\
$9.40 < \log_{10} M < 9.85$ & 840 & 2.20$_{-0.25}^{+0.26}$ & 11.61$_{-0.21}^{+0.21}$\\
$9.85 < \log_{10} M < 10.30$ & 492 & 2.46$_{-0.29}^{+0.29}$ & 11.81$_{-0.21}^{+0.21}$\\
$10.30 < \log_{10} M < 10.51$ & 163 & 2.30$_{-0.33}^{+0.33}$ & 11.69$_{-0.26}^{+0.25}$\\
$10.51 < \log_{10} M < 10.85$ & 203 & 2.61$_{-0.35}^{+0.36}$ & 11.92$_{-0.24}^{+0.25}$\\
$10.85 < \log_{10} M < 11.05$ & 97 & 2.54$_{-0.39}^{+0.37}$ & 11.86$_{-0.28}^{+0.27}$\\
\hline
\multicolumn{4}{c}{\oii~$z = 2.25$ ($\log_{10} L^\star = 42.34^{+0.04}_{-0.03}$)}\\
\hline
$42.40 < \log_{10} L_{\textrm{line}} < 42.57$ & 102 & 2.97$_{-0.42}^{+0.42}$ & 11.95$_{-0.22}^{+0.22}$\\
$42.57 < \log_{10} L_{\textrm{line}} < 43.21$ & 35 & 4.49$_{-0.73}^{+0.75}$ & 12.55$_{-0.22}^{+0.23}$\\
 & & & \\
$9.50 < \log_{10} M < 10.25$ & 61 & 3.21$_{-0.54}^{+0.66}$ & 11.95$_{-0.29}^{+0.36}$\\
$10.25 < \log_{10} M < 11.80$ & 43 & 5.01$_{-0.73}^{+0.76}$ & 12.56$_{-0.20}^{+0.21}$\\
\hline
\end{tabular}}
\label{table:OII_sample_results}
\end{table}

\section{Dependencies between Galaxy Properties and Dark Matter Halo}
\label{sec:dependencies}
In this section we present our results on how the clustering evolution of \hb~and \oii~emitters depends on line luminosities and stellar masses.

\subsection{Observed Line Luminosity Dependency}
\label{sec:line_luminosity}

\begin{figure*}
\centering
\includegraphics[width=\textwidth,trim={0in 0in 0in 0.35in},clip=true]{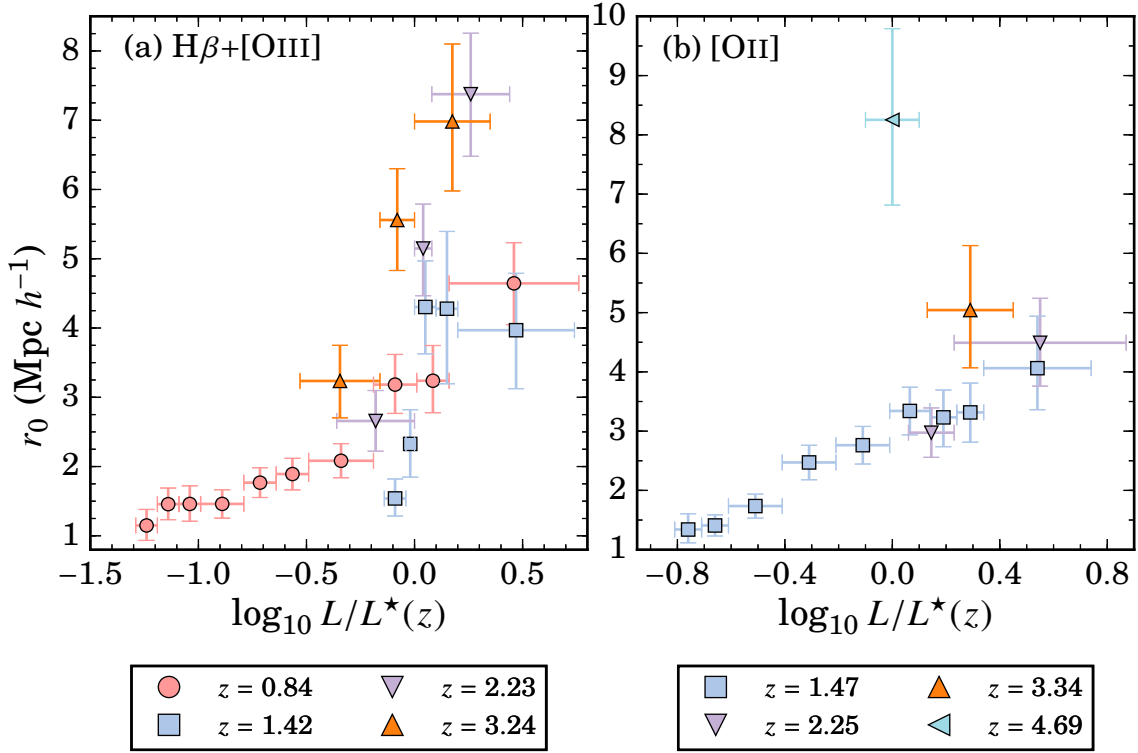}
\caption{The clustering length measured in terms of $L/L^\star(z)$. Studying the dependency of the clustering length with luminosity as a function of the ratio between line and characteristic luminosity removes the effects caused by the cosmic evolution in the luminosity functions. For each redshift slice we find that there is a strong correlation between the clustering length and $L/L^\star(z)$. There is an evolution in the clustering length such that \ro~increases with redshift at any given $L/L^\star(z)$. For example, the clustering lengths at $L \sim L^\star(z)$ are $3.2$, $4.3$, $5.2$, and $7.0~h^{-1}$ Mpc for our \hb~samples at $z = 0.84$, $1.42$, $2.23$, and $3.24$. The same strong, increasing trend between \ro~and $L/L^\star(z)$ is also seen for the \oii~sample.}
\label{fig:ro_lum}
\end{figure*}

\begin{figure*}
\centering
\includegraphics[trim={0 15 0 45},clip=true]{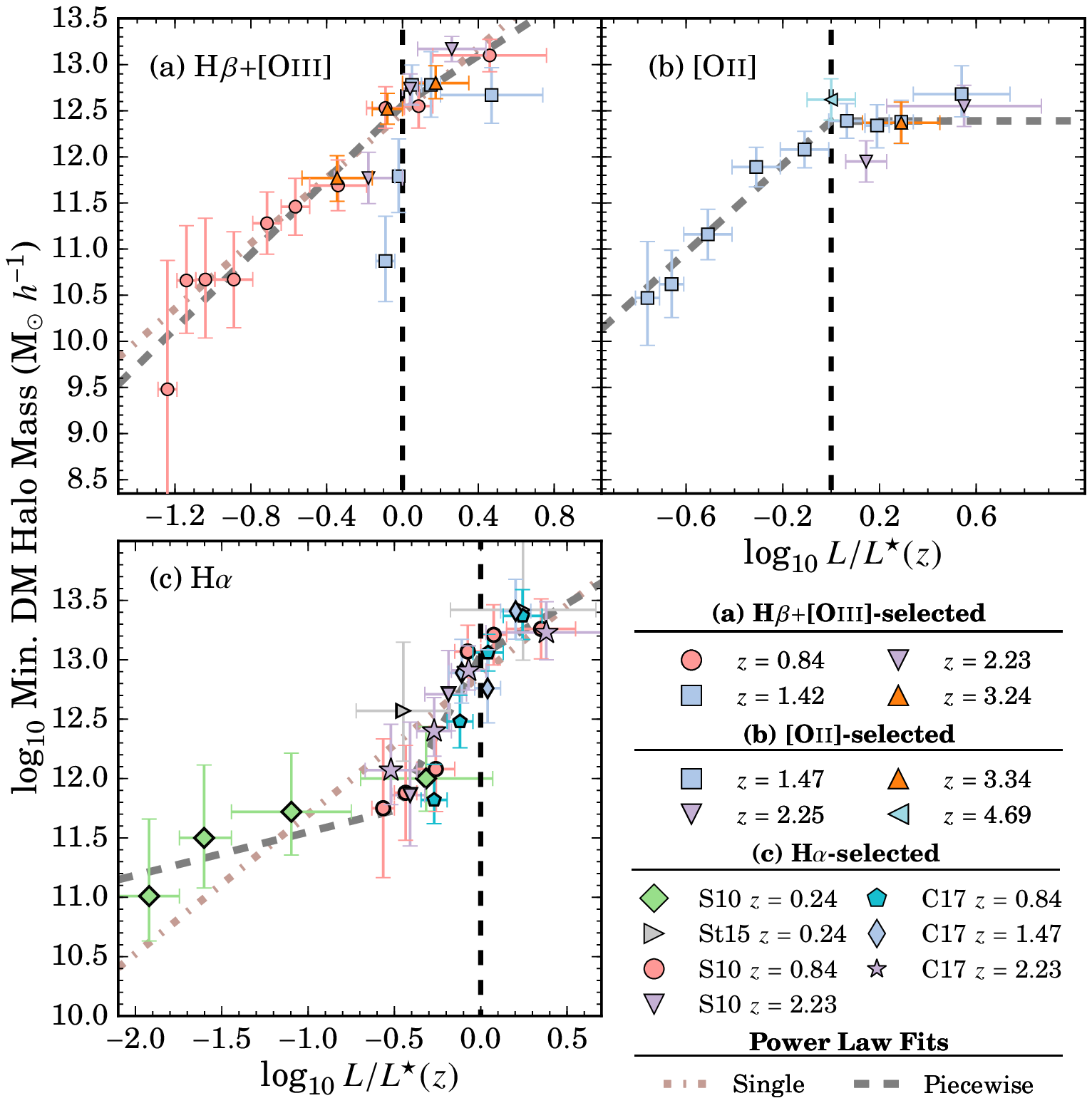}
\caption{The dependency between $L/L^\star(z)$ versus minimum halo mass for our \hb~and \oii~samples. We find a strong correlation between line luminosity and dark matter halo mass and find no redshift evolution in $L/L^\star(z)$ such that galaxies at redshifts as high as $z \sim 5$ for a given $L/L^\star(z)$ reside in halos of similar mass as galaxies at $z \sim 1$. As a comparison, we also include the \ha~measurements at $z = 0.24$ from \citet{Shioya2008} (recomputed by \citet[S10]{Sobral2010}) and \citet[St15]{Stroe2015}, $z = 0.84$ from \citet{Sobral2010}, and $z = 2.23$ from \citet{Geach2008} (recomputed by \citet{Sobral2010}). The latest \ha~results of \citet[C17]{Cochrane2017} are also included at $z = 0.84$, $1.47$, and $2.23$. The consensus from \ha~studies is a strong dependency between line luminosity and halo mass. For $L > L^\star$ emitters, we find a flat trend with halo mass consistent with $10^{12.5}$ \msol~for \oii~emitters and a shallower increasing trend for \ha~and \hb~emitters, although the scatter in the measurements are $\sim 0.5$ dex which can also be consistent with a flat trend.}
\label{fig:line_lum_dependency}
\end{figure*}

As discussed in \S \ref{sec:clustering_properties}, the clustering properties of galaxies are tied to their physical properties such that an investigation of their dependencies is required to properly map out the clustering evolution and study the connection between dark matter halos and galaxies. In this section, we study how the clustering length is dependent on the observed line luminosities and link it to the dark matter halo properties. 

Figure \ref{fig:ro_lum} shows the \ro~dependency with line luminosity normalized by the characteristic line luminosity at the corresponding redshift, $L/L^\star(z)$. The tabulated measurements are shown in Tables \ref{table:HB_sample_results} and \ref{table:OII_sample_results}. The reason we show our measurements in terms of $L/L^\star(z)$ is so that we may investigate the clustering evolution of our samples independent of the cosmic evolution of the line luminosity functions. This was motivated by the results of \citet{Sobral2010} and \citet{Cochrane2017} for their \ha~samples. \citet{Khostovan2015} showed that $L^\star(z)$ can evolve by a factor of $\sim 11 - 12$ from $z \sim 0.8 - 5$ for both \hb- and \oii-selected samples. 

For each redshift slice, we find that \ro~increases by a factor of $\sim 2 - 4$ with increasing line luminosity. There is also a redshift evolution such that at a fixed $L/L^\star(z)$, \ro~is increasing. For example, we find for our \hb~samples that the clustering length at $L \sim L^\star(z)$ is $3.2$, $4.3$, $5.2$, and $7.0~h^{-1}$ Mpc at $z = 0.84$, $1.42$, $2.23$, and $3.24$, respectively, which corresponds to a factor of $2.2$ increase in \ro~within $\sim 5$ Gyrs.

Our results suggest some redshift evolution in the clustering of galaxies as a function of line luminosity, but we must also take into account the intrinsic clustering evolution due to halos as shown in Figure \ref{fig:clustering_full_sample}. A reasonable way to assess if there is an evolution in the clustering properties is by investigating it in terms of halo masses and $L/L^\star(z)$ such that we take into account both the halo clustering (see Figure \ref{fig:clustering_full_sample}) and the line luminosity function evolutions. This relation was first studied by \citet{Sobral2010} for \ha~emitters up to $z = 2.23$ where they reported a strong, redshift-independent trend between halo mass and $L/L^\star(z)$. Here we investigate if such a relation exists for our \hb~and \oii~emitters to even higher redshifts.

Figure \ref{fig:line_lum_dependency} shows the line luminosity dependence on minimum dark matter halo masses (measured using our $r_0$-halo mass models as described in \S \ref{sec:dmh_model}) with the measurements highlighted in Tables \ref{table:HB_sample_results} and \ref{table:OII_sample_results}. We find that there is a strong relationship between line luminosity and halo mass for all redshift samples. More interestingly, we find no significant redshift evolution in the minimum dark matter halo mass such that galaxies reside in halos with similar masses independent of redshift at fixed $L/L^\star(z)$. This is found for both \hb~and \oii, as well as \ha~studies \citep{Geach2008,Shioya2008,Sobral2010,Cochrane2017}  as shown in the bottom panel of Figure \ref{fig:line_lum_dependency}. 

We quantify the observed trends by fitting both single and piecewise power laws to all measurements at all redshifts. The piecewise power laws are used in order to test the significance of a possible flattening of the observed, increasing trends for $L > L^\star(z)$. Our single power law fits are:

\begin{eqnarray}
\textrm{M}_\textrm{min} = \left\{
        \begin{array}{ll}
            10^{12.48\pm0.07} \Big(\frac{L}{L^\star(z)}\Big)^{1.77\pm0.21} & \quad \textrm{\hb} \\
            10^{12.87\pm0.06} \Big(\frac{L}{L^\star(z)}\Big)^{1.17\pm0.14} & \quad \textrm{\ha}
        \end{array}
    \right.
\label{eqn:single_power_law}
\end{eqnarray}
where we only show the measurements for \hb~and \ha~as the \oii~measurements show a clear deviation for $L > L^\star(z)$. We find that the \hb~emitters show a steeper increasing trend in comparison to \ha~but with a lower halo mass at $L\sim L^\star(z)$.

Figure \ref{fig:line_lum_dependency} shows a clear deviation from a single power law trend at $L \sim L^\star(z)$ for the \oii~samples. There is some signature of such a deviation in our \hb~and also the \ha~samples from the literature where the slope of the trends becomes shallower. We fit piecewise power laws split at $L\sim L^\star(z)$ and find:
\begin{eqnarray}
\begin{array}{l}
\textrm{\hb}:\\
\resizebox{.88\columnwidth}{!}{$
\textrm{M}_\textrm{min}= 10^{12.56\pm0.11}\left\{
        \begin{array}{ll}
            \Big(\frac{L}{L^\star(z)}\Big)^{2.02\pm0.32} &  L<L^\star \\
            \Big(\frac{L}{L^\star(z)}\Big)^{1.35\pm0.47} &  L>L^\star \\
        \end{array}
    \right.
$}
\end{array}
\label{eqn:HB}
\end{eqnarray}

\begin{eqnarray}
\begin{array}{l}
\textrm{\oii}:\\
\resizebox{.88\columnwidth}{!}{$
\textrm{M}_\textrm{min}= 10^{12.39\pm0.08}\left\{
        \begin{array}{ll}
            \Big(\frac{L}{L^\star(z)}\Big)^{2.37\pm0.31} &  L<L^\star \\
            \Big(\frac{L}{L^\star(z)}\Big)^{0.003\pm0.003} &  L>L^\star \\
        \end{array}
    \right.
$}
\end{array}
\label{eqn:OII}
\end{eqnarray}

\begin{eqnarray}
\begin{array}{l}
\textrm{\ha}:\\
\resizebox{.88\columnwidth}{!}{$
\textrm{M}_\textrm{min}= 10^{13.04\pm0.08}\left\{
        \begin{array}{ll}
            \Big(\frac{L}{L^\star(z)}\Big)^{0.36\pm0.20} &  L<0.3L^\star \\
            \Big(\frac{L}{L^\star(z)}\Big)^{2.61\pm0.36} &  0.3L^\star<L<L^\star \\
            \Big(\frac{L}{L^\star(z)}\Big)^{0.87\pm0.43} &  L>L^\star \\
        \end{array}
    \right.
$}
\end{array}
\label{eqn:HA}
\end{eqnarray}
where only the \ha~measurements includes a second split at $L \sim 0.3L^\star$ which is only constrained by the $z \sim 0.24$ \ha~measurements of \citet{Shioya2008}. Therefore, we cannot state that the trend is redshift-independent below $0.3 L^\star$ for \ha-selected emitters due to lack of measurements at different redshifts.

\begin{figure*}
\centering
\includegraphics[width=\textwidth,trim={0in 0in 0in 0.35in},clip=true]{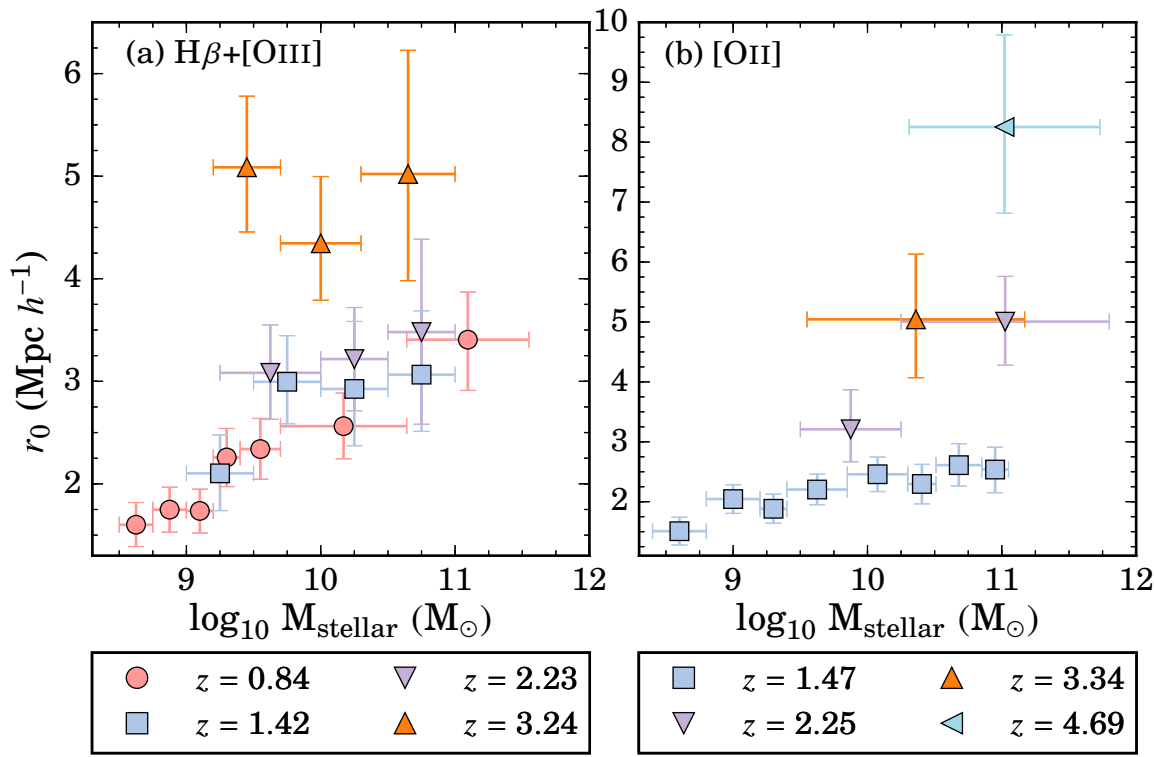}
\caption{The clustering length as measured per stellar mass bins. We find that for both \hb~and \oii~emitters the clustering length increases with increasing stellar mass. Our results show that \ro~also increases with redshift for a fixed stellar mass. In comparison to the line luminosity dependency, we find that the increasing trend with stellar mass is weaker but we note that this could be a result of the line luminosity dependency or vice versa. This is because for each stellar mass bin there is a wide range of line luminosities. We explore this inter-dependency in \S \ref{sec:grid}.}
\label{fig:ro_mass}
\end{figure*}

Equations \ref{eqn:HB} $ - $ \ref{eqn:HA} show a steep, increasing trend up to $L \sim L^\star$ followed by significantly shallower slopes beyond $L^*$. The \hb~fit shows the steepest slope of $1.35\pm0.47$ beyond $L^\star$, but we note that the spread in our halo mass measurements are quite large ($\sim 0.7$ dex) such that a flat slope can also be consistent with the measurements. The fits confirm a near constant halo mass for $L > L^\star(z)$ such that emission line-selected galaxies (\ha, \hb, and \oii) with different line luminosities $> L^\star$ reside in halos with similar masses regardless of redshift. This suggests that the mechanisms and processes causing this flattening of the line luminosity-halo mass relation is possibly the same in \ha, \hb, and \oii~emitters for all redshift slices probed. The flattening/shallower slope could also be due to the lower number density of $10^{12.5 - 13.0}$ \msol~halos given the comoving volume of our survey.

Our results also imply that there is a simple, redshift-independent relationship between the emission line luminosities of galaxies and their host halos once accounting for the evolution in $L^\star$ \citep{Sobral2010}. This has implications for theoretical studies that use photoionization codes along with semi-analytical modeling to study the connection between nebular emission lines and dark matter halo properties (e.g., \citealt{Orsi2014}).

The results reported in Equations \ref{eqn:single_power_law} $ - $ \ref{eqn:HA} and shown in Figure \ref{fig:line_lum_dependency} do not take into account the errors in $L^\star(z)$. The errors for each sample are listed in Tables \ref{table:HB_sample_results} and \ref{table:OII_sample_results}. We find that the errors are on the order of $0.05$ dex for the lowest redshift samples and $\sim 0.20$ dex for the highest redshift samples. Taking into account this error does not significantly remove the redshift independency that we see in Figure \ref{fig:line_lum_dependency}, but may change the measurements shown in Equations \ref{eqn:single_power_law} - \ref{eqn:HA}.

\subsection{Stellar Mass Dependency}
\label{sec:mass}

\begin{figure}
\centering
\includegraphics[width=1.1\columnwidth,trim={0 0 0 57},clip=true]{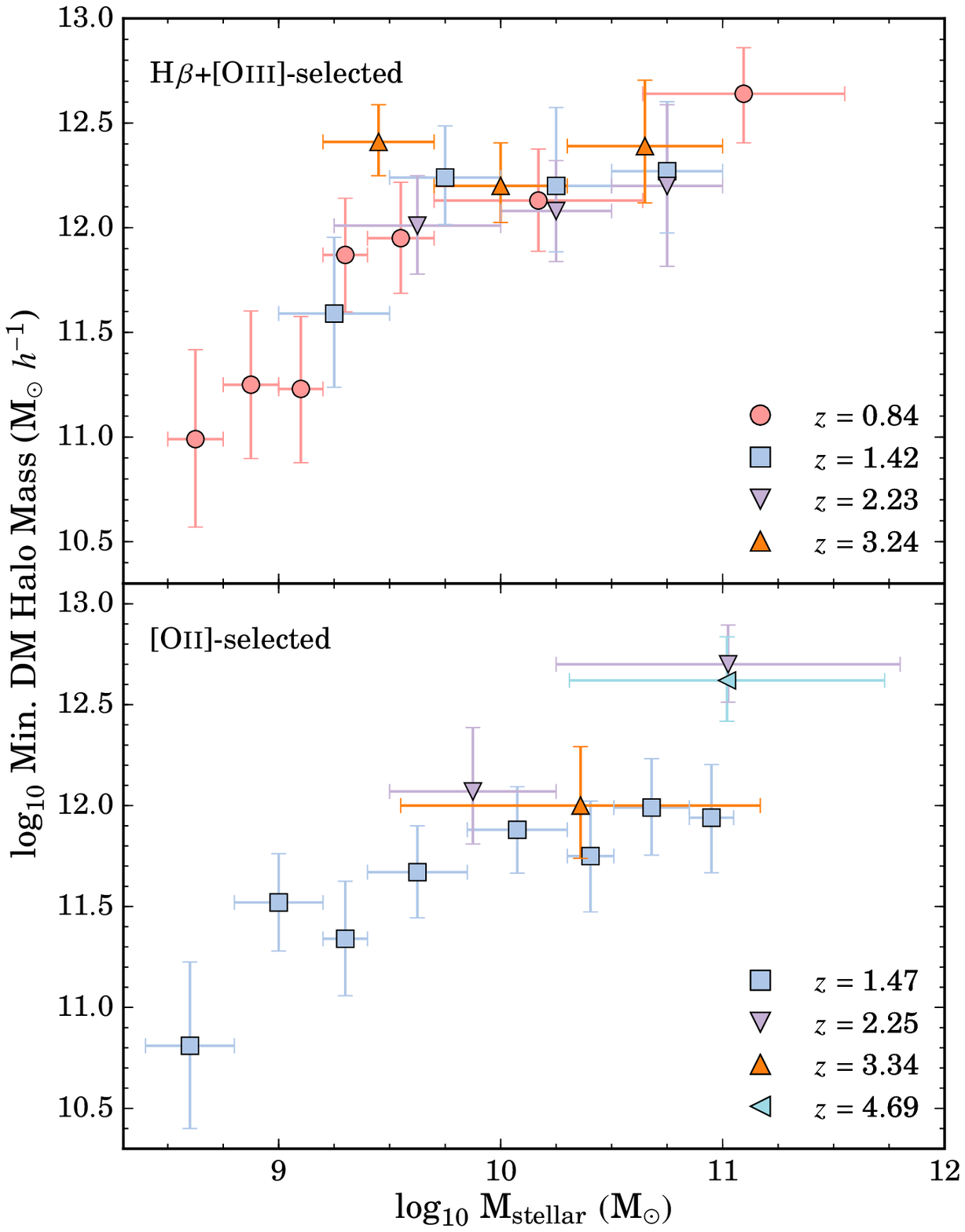}
\caption{The minimum halo mass dependency with stellar mass. We find a strong relationship at all redshift slices for our \hb~samples and for the $z = 1.47$ \oii~sample. The $z = 2.25$ \oii~sample also shows an increasing trend, but is limited only to two stellar mass bins. The other \oii~samples are limited due to sample size and could not be separated in stellar mass bins. We find no redshift evolution in the relationship. Interestingly at stellar masses $> 10^{9.75}$ \msol~the halo mass is found to be constant at $\sim 10^{12.3}$ \msol~for \hb~emitters and $\sim 10^{12}$ \msol~for \oii~emitters.}
\label{fig:mass_dependency}
\end{figure}

In principle, the mass of a halo regulates the inflow of cold gas that is used to fuel star formation activity inside galaxies with the peak in star formation activity found to occur in $10^{12}$ \msol~halos (e.g., \citealt{Peacock2000, Seljak2000,Moster2010,Behroozi2013b}). It is then expected that there is a dependency between the stellar mass of a galaxy and its host halo mass, which forms the main basis of the abundance matching technique (e.g., \citealt{Vale2004}). In this section we explore the stellar - halo mass relationship. 

Figure \ref{fig:ro_mass} shows $r_0$ per stellar mass bin for our samples of emission line-selected galaxies and listed in Tables \ref{table:HB_sample_results} and \ref{table:OII_sample_results}. Similar to the results found in \S \ref{sec:line_luminosity} for the line luminosity dependency, we find an increase in \ro~with increasing stellar mass although not as pronounced as the line luminosity dependency, especially for the high-$z$ samples. The \hb~$z = 0.84$ shows an increase of a factor of $\sim 2$ for the full range of stellar mass observed, while the $z > 1$ show an increase of a factor ranging between $1.2 - 1.5$. The \oii~$z = 1.47$ shows that the clustering length increases by a factor of $\sim 1.7$, which is weaker when compared to the line luminosity dependency. 

We also find a strong redshift evolution for a fixed stellar mass. For example, the \hb~samples show a clustering length of $2.6$, $2.9$, $3.2$, and $4.4$ Mpc $h^{-1}$ for $z = 0.84$, $1.42$, $2.23$, and $3.24$, respectively at a fixed stellar mass of $10^{10}$ \msol. To test if there is a redshift evolution we apply the same approach as was done with the line luminosity dependency by investigating the clustering evolution in terms of halo and stellar mass. We use our models as described in \S \ref{sec:dmh_model} to convert \ro~to minimum halo mass.

Figure \ref{fig:mass_dependency} shows the dependency between stellar and minimum halo mass for all redshift slices. We find a strong dependency between stellar and halo mass for $z = 0.84$ \hb~emitters between stellar masses of $10^{8.5}$ \msol~and $10^{9.8}$ \msol. There is also a hint of a dependency for $z = 1.42$ \hb~emitters in the stellar mass range of $\sim 10^{9.0}$ \msol~to $\sim 10^{10.0}$ \msol~and $z = 2.23$ \hb~emitters for a similar range, although the latter is within $1\sigma$ error bars. We find that the $z = 3.24$ \hb~sample shows a constant halo mass of $\sim 10^{12.3}$ \msol~for the full stellar mass range ($9.2 < \log_{10}$ M$_\textrm{stellar}/$\msol$ < 11.0$) and this is consistent with the other redshift slices for stellar masses $> 10^{9.75}$ \msol, although with a spread in halo mass between $12.0 < \log_{10}$ M$_\textrm{halo}/$\msol$ < 12.5$. Interestingly, we find that for the stellar mass range of $10^{9.75 - 11.0}$ \msol, \hb~emitters between $z = 0.84$ and $3.24$ reside in $\sim 10^{12.3}$ \msol~halos forming a redshift-independent plateau as also seen in the line luminosity dependency (see \S \ref{sec:line_luminosity}).

The bottom panel of Figure \ref{fig:mass_dependency} shows the dependency for \oii-selected emitters up to $z = 2.25$ and we include the full sample measurements (same as shown in Figure \ref{fig:clustering_full_sample}) for our $z = 3.34$ and $4.69$ samples due to small sample sizes. We find that the $z = 1.47$ \oii~sample shows an increase in halo mass with increasing stellar mass between $8.4 < \log_{10}$ M$_\textrm{stellar}/$\msol$ < 10$. The $z = 2.25$ sample shows an increase between $9.5 < \log_{10}$ M$_\textrm{stellar}/$\msol$ < 11.8$, although this is based only on two measurements. We find that the $z = 3.34$ measurement for the full sample is consistent with a halo mass of $\sim 10^{12}$ \msol~which agrees with the $z = 1.47$ and $2.25$ measurements. We find that the $z = 4.69$ measurements are consistent only with the most massive $z = 2.25$ \oii~emitters with a halo mass of $\sim10^{12.6}$ \msol. 

\begin{figure}
\centering
\includegraphics[width=1.1\columnwidth,trim={0 0 0 57},clip=true]{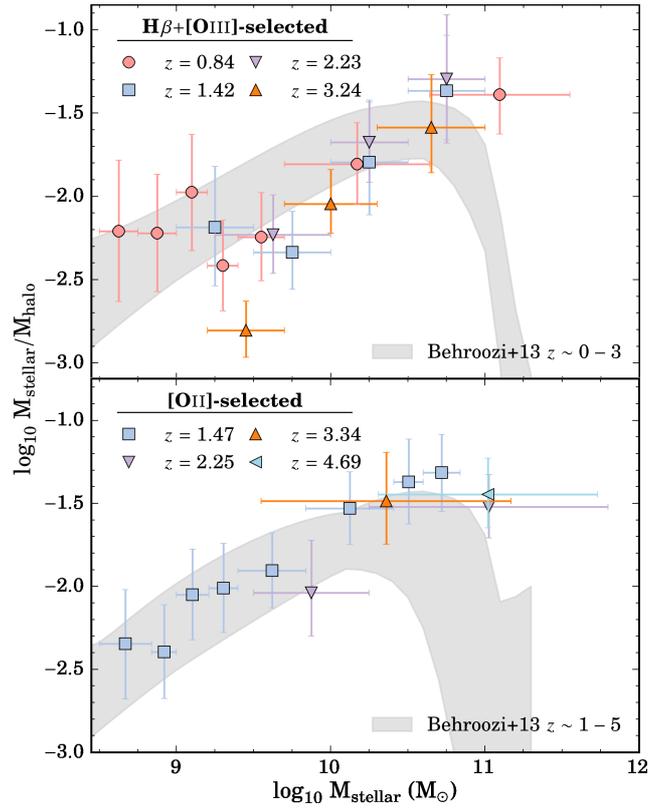}
\caption{The stellar-halo mass (SHM) ratio as a function of stellar mass. We find our \oii~measurements show a continuous, redshift-independent increase in the SHM ratio for the full stellar mass range. The \hb~measurements show a constant ratio up to $\sim 10^{9.75}$ \msol~followed by a continuous, redshift-independent increase in the ratio. We compare our measurements with the abundance matching measurements of \citet{Behroozi2013} overlaid in grey. We find that our \oii~and \hb~measurements are in agreement within $1\sigma$ except for our \hb~$10^{9.75-10.00}$ \msol~measurements.}
\label{fig:sm_hm}
\end{figure}

\subsubsection{Stellar-Halo Mass Ratio}

A byproduct of the stellar mass dependency is the stellar-halo mass (SHM, M$_\textrm{stellar}/$M$_\textrm{halo}$) ratio, which is defined as the stellar mass divided by the halo mass. This is a useful tracer of the star formation efficiency since the SHM ratio can be interpreted as the ratio of baryons that formed stars to dark matter (assuming a universal baryon fraction). Theoretical and observational studies have found that the maximum star formation efficiency in galaxies occurs in $\sim 10^{12}$ \msol~halos (e.g., \citealt{Moster2010,Moster2013,Behroozi2013b}). In this section, we explore the SHM ratio as a function of stellar mass for our \hb~and \oii~samples.

Figure \ref{fig:sm_hm} shows the SHM ratio where we find it to be redshift-independent for \hb~emitters for all stellar masses. We find the SHM ratio for $z = 0.84$ and $1.42$ \hb~emitters as constant between $8.5 < \log_{10}$ M$_\textrm{stellar}/$\msol$ < 9.75$ and increases for all redshift slices from $-2.3$ to $-1.3$ dex for M$_\textrm{stellar} > 10^{9.75}$ \msol. The bottom panel of Figure \ref{fig:sm_hm} shows the SHM ratio as a function of stellar mass for \oii~emitters. We find that the SHM ratio increases with stellar mass at $z = 1.47$ for the full stellar mass range probed. The $z = 2.25$ sample also shows an increase with stellar mass and is consistent with the $z = 1.47$ measurements. We find the same redshift-independent trend as for the \hb~emitters. 

We overlay in Figure \ref{fig:sm_hm} the measurements of \citet{Behroozi2013}, which used the abundance matching technique and the constraints set by observational measurements of the global stellar mass functions to calculate the SHM ratio up to $z \sim 8$. \citet{Behroozi2013} found that the ratio is redshift-independent and we therefore only highlight in Figure \ref{fig:sm_hm} the $1\sigma$ confidence region of their measurements that correspond to the redshifts of our sample. We find all four redshift slices for the \oii~samples are in agreement with the \citet{Behroozi2013} measurements. Our \hb~measurements are also in agreement for M$_\textrm{stellar} < 10^{9.5}$ \msol~and $> 10^{10}$ \msol. Note that the \citet{Behroozi2013} measurements are based on `global' (passive+active galaxy) stellar mass functions, while our samples are comprised of `active' galaxies (see Figure 3 of \citet{Khostovan2016} for the $UVJ$ diagram) which could explain the discrepancy at $\sim 10^{9.75}$ \msol~shown in Figure \ref{fig:sm_hm} for the \hb~samples.

\subsubsection{Minimum or Effective Halo Mass?}
\label{sec:halo_mass}

The comparison with \citet{Behroozi2013} is not exactly a like-to-like comparison as their measurements are constrained using global stellar mass functions. Our samples are emission line-selected, such that they are selecting the active population of galaxies and are not stellar mass complete. Furthermore, the halo masses reported in \citet{Behroozi2013} are defined as the mass of a host halo similar to an effective halo mass and not a minimum halo mass, as used in our work. Their models also take into account satellite galaxies, while our model assumes one central galaxy per host dark matter halo. Our measurements shown in Figure \ref{fig:sm_hm} then have two caveats: (1) stellar mass incompleteness and (2) minimum halo mass.

Despite these differences in assumptions and caveats, it is interesting that our measurements of the SHM ratio are consistent with those of \citet{Behroozi2013}. A possible reason for the agreement is that the stellar mass incompleteness and minimum halo mass effects are canceling each other. The stellar mass incompleteness could be underestimating the clustering signal and, as a consequence, underestimating the minimum halo mass. The strong agreement with our \oii~SHM ratio measurements shown in Figure \ref{fig:sm_hm} could also suggest that our \oii~samples are more representative of a stellar mass-complete sample in comparison to our \hb~samples.

In regards to the different definition of halo mass, the agreement with the measurements of \citet{Behroozi2013} could suggest that our minimum halo mass measurements are more representative of the effective halo mass in HOD models due to our simplified assumption of only pure central galaxy occupation. For example, we show the \ha~measurements of \citet{Cochrane2017} in Figure \ref{fig:line_lum_dependency} using their \ro~measurements and converting it to minimum halo mass using our model as described in \S\ref{sec:dmh_model}. We find that their effective halo masses are roughly consistent with our assessment of the minimum halo masses using their \ro~measurements. 

Since our dark matter halo model assumes only one galaxy per host halo and given the steepness of the halo mass functions, it is then likely that our minimum halo masses are similar to effective (average) halo masses. We test this by integrating the halo mass functions to calculate the effective halo mass down to a given M$_\textrm{min}$. We find that the maximum offset between the effective and minimum halo mass is $0.25$ dex at M$_\textrm{min} \sim 10^{13.5}$ \msol~and $0.07$ dex at M$_\textrm{min} \sim 10^{12}$ \msol. Our results show that even for our simplified model, the difference between minimum and effective halo mass is negligible. Although we continue to refer to our halo mass measurements as `minimum' halo mass as defined in Equation \ref{eqn:halo_bias}, we strongly caution the reader that our measurements may be more representative of the effective halo mass in comparison to HOD models.

\begin{figure}
\centering
\includegraphics[width=\columnwidth,trim={0 0 25 35},clip=true]{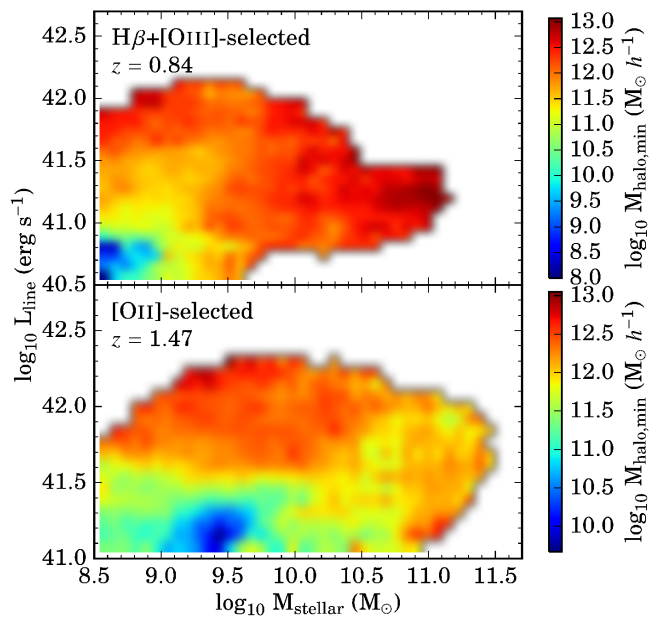}
\caption{Shown is the halo mass dependency on line luminosity and stellar mass. Only the NB921 samples are used ($z = 0.84$ \hb~and $z = 1.47$ \oii) as these are the most populated ($\sim 2500 - 3000$ sources each). All the measurements were done by randomly sampling the grid 10000 times and going through the clustering analysis to measure the halo mass. Overall, we find that the halo mass correlation with line luminosity is stronger than with stellar mass.}
\label{fig:grid}
\end{figure}

\subsection{Observed Line Luminosity -- Stellar Mass Dependency}
\label{sec:grid}
Observations have found a correlation between the star-formation rate and stellar mass in the local Universe (e.g., \citealt{Salim2007,Lee2011}), around cosmic noon (e.g., \citealt{Daddi2007,Noeske2007,Rodighiero2011,Whitaker2014,Shivaei2015}), and at higher redshifts (e.g., \citealt{Schreiber2015,Tasca2015,Tomczak2016}). Line luminosities trace star-formation activity (e.g., \oiii: \citealt{Suzuki2016}; \oii: \citealt{Kennicutt1998,Kewley2004}) and we find a dependence between halo mass, line luminosity, and stellar mass. The question that arises is how much does the dependency of line luminosity affect the dependency measured with stellar mass or vice versa?

We test this by redoing our clustering analysis in 10000 randomly selected parts of the line luminosity-stellar mass grid and calculate the halo mass following the same methodology highlighted in \S \ref{sec:clustering}. Each realization is a rectangular box randomly placed in the grid and must have $> 50$ sources. The results are shown in Figure \ref{fig:grid} for only the NB921 samples (\hb~$z = 0.84$ and \oii~$z = 1.47$) as these are the most populated samples and are much easier to investigate the dual dependency of line luminosity and stellar mass with the halo mass. We find that for increasing line luminosity and stellar mass, the halo mass is increasing from as low as $10^{8}$ to $10^{13}$ \msol, although there is a significant scatter such that to assess which property dominates the dependency with halo mass requires a look at how stellar mass (line luminosity) is dependent on halo mass for a fixed line luminosity (stellar mass). 

\subsubsection{Fixed Stellar Mass}
In this section, we investigate if there is a line luminosity dependency for a fixed stellar mass. We find a strong dependency between halo mass and line luminosity in \hb~emitters with fixed stellar masses of $10^{8.5 - 9.5}$ \msol~where the halo mass is found to increase from $\sim 10^{8.0}$ \msol~to $\sim 10^{12.5-13.0}$ \msol. Beyond $> 10^{9.5}$ \msol~the halo mass is consistent with $10^{12.5 - 13}$ \msol~for all observed line luminosities, although this is primarily due to a small sample size ($\sim 300$ sources, see Table \ref{table:HB_sample_results}) and a limiting range of line luminosities, especially at higher stellar masses.

For the $z = 1.47$ \oii~emitters, we find that for fixed stellar masses of $10^{8.5 - 11}$ \msol, there is a strong dependency with line luminosity such that the halo mass increases from $\sim 10^{9.5}$ \msol~to $10^{13}$ \msol with increasing line luminosity. Interestingly, the dependency is found for a wider range of fixed stellar masses in comparison to the \hb~sample and this could be due to the \oii~sample selecting more higher mass galaxies with low SFRs and ionization parameters compared to \hb.

\subsubsection{Fixed Line Luminosity}
In the case of a fixed line luminosity, we find that there is only a stellar mass dependency with halo mass for \hb~emitters with $L \lesssim 10^{41.5}$ erg s$^{-1}$ and it becomes more prevalent at $L \lesssim 10^{41.0}$ erg s$^{-1}$. The stellar mass dependency in the $L_\textrm{\hb} \sim 10^{41.0 - 41.5}$ erg s$^{-1}$ regime is probably due to contaminants, such as high-mass AGNs, that reside in halos of $\sim 10^{13}$ \msol. If we disregard this subpopulation of high mass sources, then the dependency breaks down. At $L_\textrm{\hb} \lesssim 10^{41.0}$ erg s$^{-1}$, we find the dependency is the strongest where emitters with stellar masses $> 10^{8.6}$ \msol~reside in increasingly higher mass halos.

Figure \ref{fig:grid} shows no significant stellar mass dependency for $z = 1.47$ \oii~emitters at a given line luminosity $> 10^{41.6}$ erg $s^{-1}$. We only find a stellar mass dependency in the case that $L_\textrm{\oii} \lesssim 10^{41.3}$ erg s$^{-1}$ where the halo mass is between $10^{11 - 11.5}$ \msol~for $8.5 < \log_{10} \textrm{M}_\textrm{stellar}/\textrm{\msol} < 9$, drops to halo masses of $10^{9.5 - 11}$ \msol~for $9 < \log_{10} \textrm{M}_\textrm{stellar}/\textrm{\msol} < 9.5$, and then increases to halo mass of $10^{12}$ \msol~with increasing stellar mass.

\subsubsection{Which one: Line Luminosity or Stellar Mass?}
We find that for both \hb~and \oii~emitters, a stellar mass dependency appears for the case of faint line luminosities as a opposed to the line luminosity dependency which appears for the full stellar mass range. This could suggest that the trend between halo mass and line luminosity are more significant than with stellar masses, such that the correlations we observed in stellar mass could be a result of the halo mass correlation with line luminosity for our samples. \citet{Sobral2010} came to a similar conclusion using a sample of $z = 0.84$ \ha~emitters and the rest-frame $K$-band luminosity as a proxy for stellar mass. \citet{Cochrane2017} also came to a similar conclusion using samples of $z = 0.84$, $1.47$, and $2.23$ \ha~emitters. We do caution the reader that our results are for line luminosity-selected samples.

\section{Discussion}
\label{sec:discussion}

In the previous sections, we found that there is a strong, redshift-independent relationship between line luminosity and minimum halo mass (relatively independent of stellar mass for $z = 0.84$ and $1.47$ \hb~and \oii~emitters, respectively) up to $L \sim L^\star$ for \ha, \hb, and \oii~emitters. For the $L > L^\star$ regime, we find that the dependency becomes shallower and is consistent with minimum halo masses between $10^{12.5}$ \msol~and $10^{13}$ \msol. In this section, we discuss potential physical reasons for the flattening/shallower slope of this relationship for the brightest emitters with the understanding that the emission lines observed trace the underlying star formation activity.

\subsection{Transitional Halo Mass}
Current models of galaxy formation suggest that the star formation efficiency is tied to the host halo mass with the peak in the efficiency found in $\sim 10^{12}$ \msol~halos (e.g., \citealt{Behroozi2013b}). For $> 10^{12}$ \msol~halos, models predict that the star formation activity in galaxies diminishes as external quenching mechanisms (e.g., shock heating of infalling gas; \citealt{Dekel2006}) become stronger and are accompanied by internal quenching mechanisms (e.g., AGN feedback; \citealt{Best2006}). This is referred to as `halo quenching', where a specific global halo mass is related to the quenching of galaxies. We note that this is still debatable where, observationally, some studies have found that external quenching is mainly a local phenomenon (e.g., \citealt{Darvish2016}) and does not depend significantly on the global halo mass (e.g., \citealt{Peng2012, Carollo2013}). Other observational studies find that galaxy quenching does depend on halo mass (e.g., \citealt{Prescott2011}; also see references in \citealt{Darvish2017}).

A consequence of the halo quenching predictions is a possible characteristic halo mass scale for which the fraction of star-forming galaxies drops and the fraction of passive galaxies increases sharply. Current predictions are that this occurs around a few $\times~10^{12}$ \msol~to $10^{13}$ \msol~and is also redshift independent \citep{Croton2006,Dekel2006,Cen2011,Bower2017}. Observations have reported such a transitional halo mass. For example, \citet{Dolley2014} used a sample of $\sim 23,000$ 24\micron-selected sources between $0.2 < z < 1.0$ with an areal coverage of $8.42$ deg$^2$ and find evidence for a transitional halo mass $\sim 8 \times 10^{12}$ \msol. \citet{Hartley2013} came to a similar conclusion using the deep $0.77$ deg$^2$ UKIDSS UDS data up to $z \sim 3$ and measure a transitional halo mass of $5 \times 10^{12}$ \msol.

We show in Figure \ref{fig:line_lum_dependency} that $L > L^\star$ emitters have a flat/shallower line luminosity dependency consistent with minimum halo masses between $3\times10^{12}$ \msol~and $10^{13}$ \msol. Note that based on our short discussion in \S \ref{sec:halo_mass}, our minimum halo masses may be more representative of the effective halo mass due to our assumptions made in \S \ref{sec:dmh_model}. With this caveat taken into account, our results are then consistent up to $z \sim 5$ with the predictions of a transitional halo mass for which the number of star-forming galaxies (traced by our sample) diminishes and the fraction of passive galaxies increases. We note that this can also be a sample selection effect since the number densities of $> 10^{13}$ \msol~halos decreases significantly and requires large comoving volumes to detect their residing galaxies. 

\subsubsection{Potential Causes}
Although our results show evidence for this transitional halo mass, it raises the question of how the brightest emitters reside in $10^{13}$ \msol~halos. Since the line luminosity traces the star-formation activity, it then seems puzzling that systems with such high SFRs are found in massive halos when the peak SF efficiency is found in $\sim 10^{12}$ \msol~halos. One possibility is that $L > L^\star$ emitters have their emission lines powered by AGN activity rather than SF activity. \citet{Sobral2016} spectroscopically followed up 59 bright $L > L^\star$ \ha~emitters and found that the AGN fraction increases with observed line luminosity such that the fraction of AGNs is $\sim 50 \%$ by $\sim 4 L^\star$ . Although this is only measured for \ha~emitters up to $z = 2.23$ and may not be true for \hb~and \oii~emitters, studies up to $z \sim 1.6$ have shown that X-ray and radio-selected AGN tend to reside in halos of $\sim 10^{13}$ \msol~\citep{Hickox2009, Koutoulidis2013, Mendez2016}, which is consistent with the constant halo mass for $L > L^\star$ \hb~and \oii~emitters shown in Figure \ref{fig:line_lum_dependency}. Therefore, it is quite possible that these sources are AGN, although we require spectroscopic follow-up to measure AGN fractions in this line luminosity range for the \hb~and \oii~samples.

Another possibility is that a fraction of the brightest emitters can have their emission lines powered by major merging events, such that these systems are currently undergoing a starburst phase. Simulations of major mergers predict elevated levels of star-formation activity (e.g., \citealt{Mihos1996,diMatteo2008,Bournaud2011}) and observations have thus far found evidence to support this (e.g., \citealt{Hung2013}). Semi-analyical models have also predicted that the stellar mass assembly in high-mass halos is due to mergers (e.g., \citealt{Zehavi2012}). A detailed morphological study of the fraction of mergers as a function of line luminosity would help in addressing this issue and we plan to explore this in the future.

It could also be possible that environmental effects could allow for the presence of $> L^\star$ emitters in massive halos. \citet{Dekel2006} used simulations and predict that cold filamentary streams can penetrate the shock heated halo gas in $> 3\times 10^{12}$ \msol~halos and fuel star-formation activity in $L > L^\star$ galaxies above $z > 2$. To support this level of star-formation activity requires large cold gas accretion rates and a recent ALMA study by \citet{Scoville2017} estimated the rate to be $> 100$ \msol~yr$^{-1}$ for $z > 2$ to maintain galaxies along the main-sequence.

Overall, we find evidence for a possible transitional halo mass for which star-forming galaxies become less common and halos are increasingly populated by passive galaxies. It stands to reason that the $L > L^\star$ emitters are a mixture of AGN- and star-formation-dominated systems. This is also suggested by \citet{Kauffmann2003} in the local Universe (up to $z \sim 0.3$) where they find that galaxies with AGN and bright \oiii~lines also include young stellar populations due to a recent phase of star-formation activity. Future spectroscopic and morphological studies can shed light on the physical processes involved that are powering nebular emission lines in such massive halos and provide us with valuable insight on the quenching mechanisms that are occurring at this transitional halo mass.

\subsection{Clustering more dependent on line luminosity than stellar mass?}
In \S\ref{sec:mass} and \S\ref{sec:grid} we found that the dependency of clustering on line luminosity was more significant than on stellar mass. We also concluded, based on the results of our $z = 0.84$ \hb~and $z = 1.47$ \oii~samples in \S\ref{sec:grid}, the stellar mass dependency may be a result of the line luminosity dependency. This is a similar conclusion made by \citet{Sobral2010} where they used a $z = 0.84$ \ha-selected sample and found that the line luminosity dependency was more significant than the dependency with stellar mass. \citet{Coil2017} came to a similar conclusion where they found that the clustering amplitude was a stronger function of the specific star formation rate than stellar mass and that the clustering strength for a given specific star formation rate was found to be independent of stellar mass. \citet{Cochrane2017} used \ha-selected narrowband samples at $z = 0.84$, $1.47$, and $2.23$ and found that the line luminosity dependency was not driven/independent of stellar mass.

We note that the lack of a strong stellar mass dependency with clustering strength/dark matter halo mass could be mainly caused by sample selection. As mentioned before, our samples are selected based on line flux such that they are complete in line luminosity down to a completeness limit. In regards to stellar mass, our samples are not complete, especially for the low stellar mass range ($< 10^9$ \msol; see \citealt{Khostovan2016} for the stellar mass functions of our samples). We can only conclude that for narrowband-selected samples, the clustering strength dependency with stellar mass seems to be less significant than the dependency with line luminosity and may also be a result of it as well.

\section{Conclusions}
\label{sec:conclusions}

We have presented our \hb~and \oii~clustering measurements up to $z \sim 3.3$ and $\sim 4.7$, respectively. The main results of this study are:

\begin{enumerate}[(i)]
\item We find that the power law slopes of the angular correlation functions are consistent with $\beta \sim -0.80$. Using the exact Limber equation, we find typical \ro~between $1.45 - 4.01$ $h^{-1}$ Mpc and $1.99 - 8.25$ $h^{-1}$ Mpc for \hb~and \oii~emitters, respectively. These correspond to minimum halo masses between $10^{10.70 - 12.08}$ \msol~and $10^{11.46 - 12.62}$ \msol, respectively.

\item A \ro-line luminosity dependency is found where the brightest emitters are more clustered compared to the faintest emitters. This dependency is found to be redshift-dependent but is biased due to evolution in the line luminosity function. When rescaling based on $L^\star(z)$ and using model predictions of halo mass given \ro, we find a strong increasing dependency between minimum halo mass and line luminosity that is independent of redshift with the faintest \hb~(\oii) emitters found in $10^{9.5}$ \msol~($10^{10.5}$ \msol) halos and the brightest \hb~(\oii) emitters in $10^{13}$ \msol~($10^{12.5}$ \msol) halos. 

\item A stellar mass dependency trend is found with \ro~and, when converted to minimum halo mass, is found to be redshift independent. We find that \hb~emitters with stellar masses $> 10^{9.75}$ \msol~reside in $10^{12.3}$ \msol~halos between $z = 0.84$ and $3.24$. The \oii~samples also show a stellar mass dependency for the full stellar mass range.

\item We investigate how the interrelation between observed line luminosity and stellar mass can affect the individual dependencies we see on minimum halo mass. By creating subsamples in a line luminosity-stellar mass grid space for the most populated samples (\hb~$z = 0.84$ and \oii~$z = 1.47$), we find that the main dependency on minimum halo mass arises from the observed line luminosity such that the stellar mass dependency is weaker and could be a result of the line luminosity dependency. This then suggests a simple connection between the nebular emission line properties of galaxies and their host halo mass.

\item The line luminosity-halo mass dependency shows an increase from the faintest emitters observed to $L \sim L^\star(z)$. For emitters brighter that $L^\star$, we find that the trend is consistent with halos between $10^{12.5 - 13}$ \msol. To understand what is powering such bright emission lines, we consider three possibilities: AGN-driven, merger-driven, and/or gas inflow. There is evidence from related studies to support this hypothesis although spectroscopic and morphological studies of our samples are required to properly investigate these sources. In comparison to predictions from models, we find that the shallower trend that we observe for $L > L^\star(z)$ emitters is consistent with the transitional halo mass for which the fraction of star-forming galaxies decreases and the fraction of passive galaxies increases due to internal and external quenching mechanisms.

\end{enumerate}

Our results suggest a simple connection between the clustering/dark matter halo properties and nebular emission line properties of star-forming/`active' galaxies up to $z \sim 5$. This has implications for future theoretical studies that model this connection since previous constraints were up to $z \sim 2$ for only \ha~emitters.  On the observational side, future spectroscopic studies of bright, emission line-selected galaxies can allow us to investigate the dependency between the ISM properties (internal mechanisms) of galaxies and massive halos (external mechanisms). Morphological studies of our samples can also test to see if the shape of galaxies is connected with the host halo properties. Future space-based (e.g., {\it JWST}, {\it WFIRST}) and ground-based observatories (e.g., European Extremely Large Telescope, Thirty Meter Telescope), can also allow us to study the clustering properties of emission line-selected galaxies at higher redshifts and larger comoving volumes. This would allow us to see when the following redshift-independent trends that seem to have been in place since $z \sim 5$ were first established, which would present a new scaling relation for galaxy formation and evolution models.

\section*{Acknowledgments}
AAK thanks Anahita Alavi and Irene Shivaei for useful discussion in the making of this paper. 

AAK acknowledges that this work was supported by NASA Headquarters under the NASA Earth and Space Science Fellowship Program - Grant NNX16AO92H.
DS acknowledges financial support from the Netherlands Organisation for Scientific Research (NWO) through a Veni fellowship and from Lancaster University through an Early Career Internal Grant A100679. PNB is grateful for support from STFC via grant STM001229/1. IRS acknowledges support from STFC (ST/L00075X/1), the ERC Advanced Grant DUSTYGAL (321334), and a Royal Society/Wolfson Merit award. JM acknowledges the support of a Huygens PhD fellowship from Leiden University. BD acknowledges financial support from NASA through the Astrophysics Data Analysis Program (ADAP), grant number NNX12AE20G.

\bibliography{clustering_ELGs}

\appendix
\section{Where does Limber's Approximation fail?}
\label{sec:limber_fails}
As discussed in \S \ref{sec:real_space}, Limber's approximation works well up to a certain angular separation. The question that arises is to what angular scales can the Limber approximation be used? As \citet{Simon2007} showed, this depends on the filter profile. Here, we briefly describe the method to calculate the angular scales for which Limber's approximation fails. The equations derived in \citet{Simon2007} assumed a simple top-hat filter.

The Limber equation is generally defined as:
\begin{eqnarray}
w(\theta) & \approx & \int_0^\infty \mathrm{d}r_1 \int_0^\infty \mathrm{d}r_2~p_1(r) p_2(r) \xi\Big(R,\frac{r_1 + r_2}{2}\Big) \nonumber \\
R & = & \sqrt{r_1^2 + r_2^2 - 2 r_1 r_2 \cos{\theta}}
\label{eqn:limber_exact}
\end{eqnarray}
where $p(r)$ is the filter profile defined by a center, $r_c$, and width, $\Delta r$, in comoving distance units. The comoving distance between the observer and source 1 and 2 are defined as $r_1$ and $r_2$ with $R$ being the radial separation using the law of cosines. \citet{Simon2007} showed that in the case that $r_c \gg \Delta r$ the ACF becomes a rescaled version of the spatial correlation function with a slope of $\gamma$ instead of $1-\gamma$:
\begin{eqnarray}
w(\theta_b) \approx \xi(r_c \theta_b, r_c)
\label{eqn:rescaled}
\end{eqnarray}
where we define $\theta_b$ as the angular separation for which the Limber approximation fails. Note that the $r_c \theta_b$ term arises from doing a small-angle approximation in the definition of $R$ (see Equation \ref{eqn:limber_exact} with $r_1 = r_2 = r_c$). 

To calculate $\theta_b$, we rewrite Equation \ref{eqn:rescaled} using the definition of $A_w$ in the Limber approximation to get:
\begin{eqnarray}
\theta_b &=& \frac{1}{A_w} \Bigg(\frac{r_0}{r_c}\Bigg)^\gamma\\
 &=& \frac{1}{r_c^\gamma \sqrt{\pi}} \frac{\Gamma(\gamma/2)}{\Gamma{\Big(\frac{\gamma-1}{2}\Big)}} \Bigg[\int_0^\infty \mathrm{d}\bar{r} p_1(\bar{r}) p_2(\bar{r}) \bar{r}^{1-\gamma} \Bigg]^{-1}.
\end{eqnarray}
where $\bar{r}$ is defined as $(r_1+r_2)/2$ and $\Gamma$ is the gamma function. The above equation is generalized that for any filter $p(\bar{r})$, the expected angular separation for which the Limber approximation departs from the true ACF can be measured. In this study, we treated all four narrowband filters as Gaussians and show the true and Gaussian filter profile parameters in Table \ref{table:gauss_filter_params}. We find that all our narrowband filters can be well-treated as Gaussian filters. 

% TABLE: TRUE AND GAUSSIAN FILTER PROFILE PARAMETERS
\begin{table}
\centering
\resizebox{1.0\columnwidth}{!}{
\begin{tabular}{ccccccc}
\hline
\multicolumn{7}{c}{True \& Gaussian Filter Profile Parameters}\\
\hline
Filter & $\lambda_\textrm{obs}$ & \textrm{FWHM} & $z_\textrm{True}^{\textrm{\hb}}$ & $z_\textrm{Gaussian}^{\textrm{\hb}}$ & $z_\textrm{True}^{\textrm{\oii}}$ & $z_\textrm{Gaussian}^{\textrm{\oii}}$\\
          &    (\micron)                      &      (\AA)        &   & & &\\
 \hline
 NB921 & 0.9196 & 132 & $0.84\pm0.01$ & $0.83\pm0.01$ & $1.47\pm0.02$ & $1.46\pm0.01$\\
 NBJ	     & 1.211   & 150 & $1.42\pm0.02$ & $1.42\pm0.01$ & $2.25\pm0.02$ & $2.25\pm0.02$\\
 NBH     & 1.617   & 211 & $2.23\pm0.02$ & $2.23\pm0.02$ & $3.34\pm0.03$ & $3.34\pm0.02$\\
 NBK     &  2.121  & 210 & $3.24\pm0.02$ & $3.24\pm0.02$ & $4.69\pm0.03$ & $4.69\pm 0.02$\\
 \hline
 \end{tabular}}
 \caption{The True filter parameters and the corresponding Gaussian-assumed filter parameters in terms of the \hb~and \oii~redshifts.}
 \label{table:gauss_filter_params}
 \end{table}

\section{Uncertainties in Clustering Length - Dark Matter Halo mass predictions}
\label{sec:halo_uncertain}

We presented our \ro-halo mass predictions in \S \ref{sec:dmh_model} where we assumed a \citet{Tinker2010} halo bias function and \citet{Tinker2008} halo mass function. Other prescriptions for the mass and bias functions do exist and in this section we explore how our predicted minimum halo masses change when assuming different assumptions. We consider three cases. The {\it first} case is assuming the \citet{Press1974} mass function and \citet{Mo1996} bias function. This is a `classical' assumption and was also used in \citet{Matarrese1997} and \citet{Moscardini1998}, from which we use their methodology to make our predictions as highlighted in \S \ref{sec:dmh_model}. The {\it second} case assumes the \citet{Sheth2001} halo mass and bias functions. Lastly, the {\it third} case assumes the \citet{Tinker2008} halo mass function and \citet{Jose2016} bias function. 

\begin{figure}
\centering
\includegraphics[width=\columnwidth,trim= {10 2 43 25},clip=true]{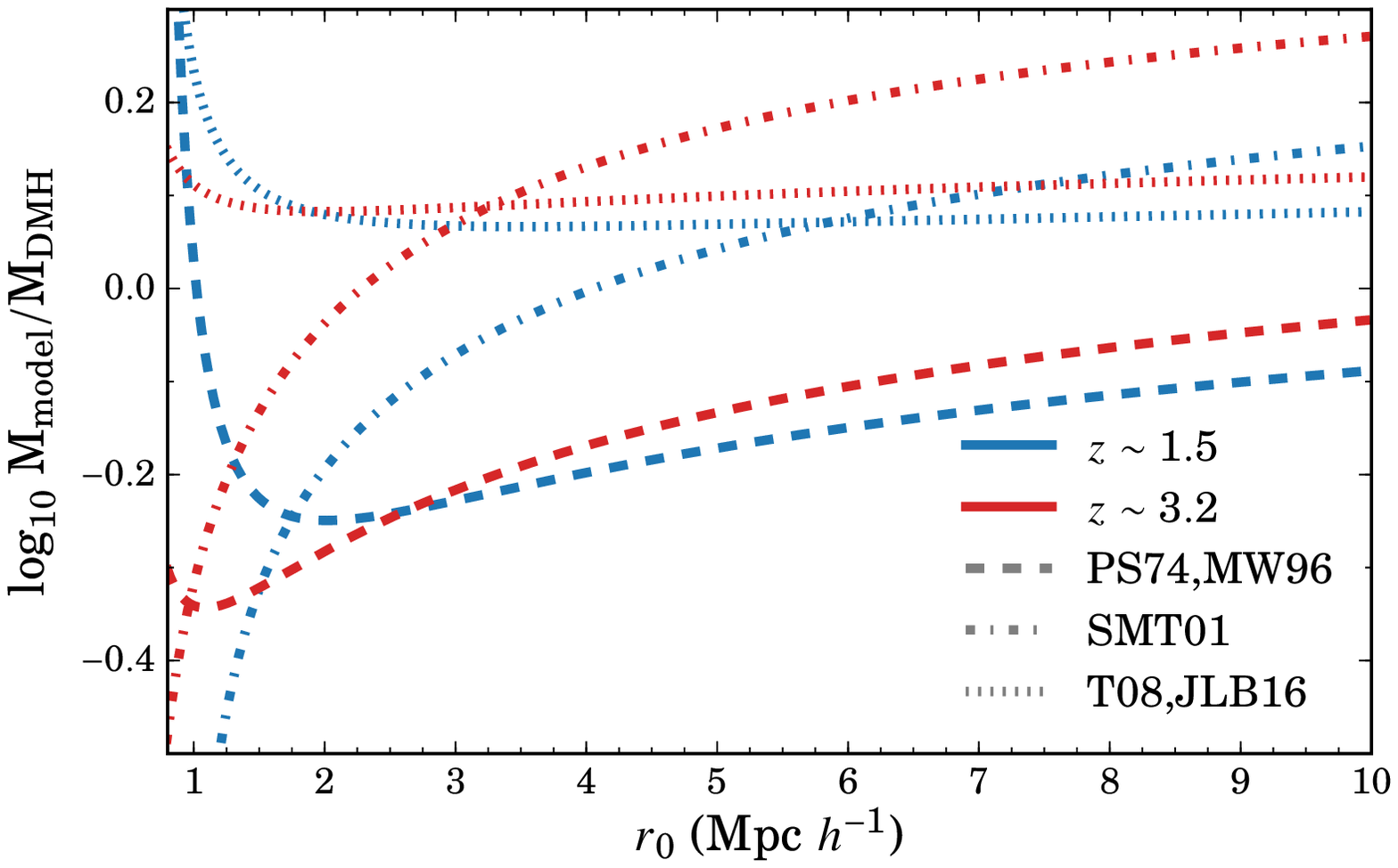}
\caption{Comparison of the predicted minimum halo masses for a given \ro~between our assumed halo bias \citep{Tinker2010} and mass \citep{Tinker2008} functions against other assumptions. The cases are as follows: (1) \citet{Press1974} mass function and \citet{Mo1996} bias function, (2) \citet{Sheth2001} mass and bias functions, and (3) \citet{Tinker2008} mass function and \citet{Jose2016} bias function. We show the difference for $z \sim 1.5$ and $\sim 3.2$ with M$_\textrm{model}/$M$_\textrm{DMH}$ being the ratio of one of the cases highlighted above (M$_\textrm{model}$) and our model (M$_\textrm{DMH}$) as described in \S \ref{sec:dmh_model}. We find that assuming different prescriptions for halo properties can introduce offsets of $\sim \pm0.2$ dex for \ro$>3 $Mpc $h^{-1}$ and become significantly worse for \ro$<3$ Mpc $h^{-1}$ such that at \ro$\sim 1$ Mpc $h^{-1}$ the offset range between $-0.4$ to $0.2$ dex.}
\label{fig:errors}
\end{figure}

Figure \ref{fig:errors} shows a comparison between our predictions of halo mass (M$_\textrm{DMH}$) and the predictions from the three cases highlighted above (M$_\textrm{model}$) for a given \ro~measurement at $z \sim 1.5$ and $\sim 3.2$. We find that offsets can be quite large, especially towards lower \ro~values where above 3 Mpc $h^{-1}$ we find offsets of $\pm 0.2$ dex and below 3 Mpc $h^{-1}$ the offsets can be as high as 0.4 dex. The only case that best matches our predictions is the third case, which is not surprising as it uses the \citet{Tinker2008} halo mass function (same as the one we assumed) and the \citet{Jose2016} bias function is an update of the \citet{Tinker2010} bias function. Based on Figure \ref{fig:errors}, we caution the reader that minimum halo mass measurements, be it from our model in \S \ref{sec:dmh_model} or any HOD/abundance matching model, are sensitive to the assumed halo prescriptions.

% Don't change these lines
\bsp	% typesetting comment
\label{lastpage}
\end{document}